\RequirePackage{rotating}

\documentclass[a4paper,fleqn,usenatbib]{mnras}

\usepackage{newtxtext,newtxmath}

\usepackage{graphicx}

\newcommand {\be} {\begin{equation}}
\newcommand {\ee} {\end{equation}}

\usepackage[normalem]{ulem}
\usepackage{color}
\definecolor{KR}{RGB}{250, 0, 0}	
\definecolor{nKR}{RGB}{130,130,130}	

\usepackage{upgreek}	
\usepackage{threeparttablex}
\usepackage{lscape}
\usepackage{afterpage}
\usepackage{capt-of}
\usepackage{lipsum}
\usepackage{subcaption}
\usepackage{rotating}
\usepackage[T1]{fontenc}
\usepackage{ae,aecompl}

\usepackage{graphicx}	

\title[X-ray spectra of radio galaxies and RQ counterparts]{Comparison of hard X-ray spectra of luminous radio galaxies and their radio-quiet counterparts}

\author[Gupta et al.]{
Maitrayee~Gupta,$^{1}$\thanks{E-mail: mgupta@camk.edu.pl}
Marek~Sikora,$^{1}$
Katarzyna~Rusinek$^{1}$
and Greg~M.~Madejski$^{2}$
\\
$^{1}$Nicolaus Copernicus Astronomical Center, Bartycka 18, 00-716 Warsaw, Poland\\
$^{2}$Kavli Institute for Particle Astrophysics and Cosmology, Stanford University, Stanford, CA 94305, USA
}

\date{Accepted XXX. Received YYY; in original form ZZZ}

\pubyear{2015}

\begin{document}
\label{firstpage}
\pagerange{\pageref{firstpage}--\pageref{lastpage}}
\maketitle

\begin{abstract}

We study the differences in X-ray properties of luminous radio galaxies and their radio-quiet counterparts. In order to avoid biases associated with the dependence of X-ray properties on the black hole mass and Eddington ratio, the radio-loud and radio-quiet objects are selected to cover similar ranges of these parameters.
Our studies are based on the X-ray data from the {\it Swift}/BAT catalogue. We confirm previous results that radio galaxies are on average X-ray-louder than radio-quiet AGNs, but find that their spectral slopes are very similar.
This suggests that in radio-loud and radio-quiet AGNs the hard X-rays are produced in the same region and by the same mechanism. We argue that this region can be associated with the hot, geometrically thick, central portion of the accretion flow, where production of hard X-rays is likely to be dominated by Comptonization of the optical/UV radiation of the truncated 'cold' accretion disc by hot electrons.  The larger X-ray luminosities in radio-loud AGNs may result from larger radiative efficiencies of the innermost portions of the accretion flows around faster rotating black holes.
 
\end{abstract}

\begin{keywords}
quasars --- galaxies: jets --- radiation mechanisms: non-thermal --- acceleration of particles
\end{keywords}



\section{Introduction}

 Known already since the 1980s, the so-called radio-dichotomy
  of AGNs \citep{1980A&A....88L..12S,1986MNRAS.218..265P,1989AJ.....98.1195K,1990MNRAS.244..207M}
  is still awaiting theoretical explanation.  Their radio loudness parameter, defined as the ratio of
  radio luminosity of  associated with them radio sources to optical luminosity of their accretion discs,
  spans  several orders of magnitude (e.g. \citealt{2002AJ....124.2364I,2007ApJ...654...99W,2012ApJ...759...30B,2015AJ....149...61K}),
  and this implies similarly  large range of the jet production efficiency (e.g. \citealt{2013ApJ...764L..24S}). 
  Most luminous radio sources need to be powered by jets at rates approaching or sometimes even
  exceeding accretion luminosities (e.g. \citealt{1991Natur.349..138R,2007MNRAS.374L..10P,2011MNRAS.411.1909F,2017MNRAS.466.2294R}).
  According to recent theoretical and numerical studies, the production of such powerful jets can be via
  the Blandford-Znajek mechanism \citep{1977MNRAS.179..433B} involving a fast rotating black hole immersed in very strong
  magnetic field.  Such field is found to be maximized in the model called magnetically-arrested-disc
  (MAD) scenario, in which magnetic flux is confined on the BH by ram pressure of the accretion flow
  \citep{2003PASJ...55L..69N,2008ApJ...677..317I,2009ApJ...704.1065P,2011MNRAS.418L..79T,2012MNRAS.423.3083M}.
  If that is true, one might expect that structure and physics  of accretion flows in such objects may differ
  significantly from those in radio-weak AGNs.  However their averaged SEDs (spectral energy distributions) in IR, 
  optical and UV bands are very similar
  \citep{1994ApJS...95....1E,2006AJ....131..666D,2006ApJS..166..470R,2011ApJS..196....2S,2016MNRAS.461.2346G},
  and statistically significant differences are found only in X-ray bands. Specifically, RL quasars have been
  found to be X-ray-louder and to have harder X-ray spectra  than RQ quasars
  \citep{1981ApJ...245..357Z,1987ApJ...313..596W,1987ApJ...323..243W}.  
  These results were confirmed also for lower-redshift type 1 AGNs represented 
by broad-line radio galaxies (BLRGs) and radio-quiet (RQ) Seyfert 1 galaxies 
\citep{1998MNRAS.299..449W,1999ApJ...526...60S,2006ApJ...642..113G,2011ApJ...740...29K,2011ApJ...726...20M}.  
In addition, it was found that X-ray reflection features are weaker
in radio-loud (RL) than in RQ AGNs, and this led to speculations that
it is so because in RL AGNs the X-ray emission is contributed not only by 
the disc corona, as originally proposed by \citet{1979MNRAS.189..421L} and \citet{1991ApJ...380L..51H}, but also by the base of a jet (e.g. \citealt{1998MNRAS.299..449W}).  
  An alternative possibility was that accretion discs in luminous RL AGNs could be on average more 
  ionized than in Seyferts \citep{2002MNRAS.332L..45B,2010ApJ...710..859E,2014ApJ...794...62B} 
  or that geometrically thin and optically thick accretion discs in such AGNs are truncated at some
  radius, and that X-rays are emitted by the hot ADAF (Advection Dominated Accretion Flow) formed
  within that radius (e.g. \citealt{1994ApJ...428L..13N,2014ARA&A..52..529Y}). The latter possibility
  was specifically explored following detailed observations of individual BLRGs in different X-ray bands 
  (e.g. \citealt{2000ApJ...537..654E,2007ApJ...659..235G,2009ApJ...700.1473S,2010ApJ...721.1340T,2012ApJ...752L..21C,2013ApJ...772...83L}).
  However, all those comparisons of X-ray properties of RL and RQ AGNs were performed selecting them using different methods and/or having their samples 
  with very different  average  Eddington ratios (usually significantly lower in RL AGNs than in RQ ones) and BH
  masses (typically much larger in RL AGNs than in RQ ones, e.g. \citealt{2004MNRAS.353L..45M,2011MNRAS.416..917C,2017ApJ...846...42K}).

Currently, with much larger samples of AGNs detected in the hard X-ray band 
as provided by the {\it Swift}/BAT all-sky surveys \citep{2013ApJS..207...19B,2018ApJS..235....4O}
it became possible to study the differences between X-ray spectra 
in RL and RQ AGNs using their subsamples covering similar
ranges of BH masses and Eddington ratios. 
Furthermore, by using the mid-IR rather than optical data to estimate 
AGN bolometric luminosities, we expanded the compared samples by including
both type 1 and type 2 AGNs, but reduced them to cover the ranges of BH masses 
and Eddington ratios typical for luminous FR\,II radio galaxies \citep{1974MNRAS.167P..31F}.
The resulting subsamples are represented
by 24 broad-line and narrow-line radio galaxies, and by 46 broad-line 
and narrow-line radio-quiet Seyfert galaxies.
 
The paper is organized as follows:  
in Sections 2 and 3 we briefly describe our selection procedure and data 
processing. In Section 4 we present the results of our comparison analysis.  
In Section 5 we provide possible interpretation of these results and  in 
Section 6 we summarize our findings.

Throughout the paper we assume a $\Lambda$CDM cosmology with 
$H_0 = 70 \ {\rm km} \ {\rm s}^{-1} \ {\rm Mpc}^{-1}$, $\Omega_m=0.3$,
and $\Omega_{\Lambda}=0.70$.  
 
\begin{figure*}
\centering
\includegraphics[width=\textwidth]{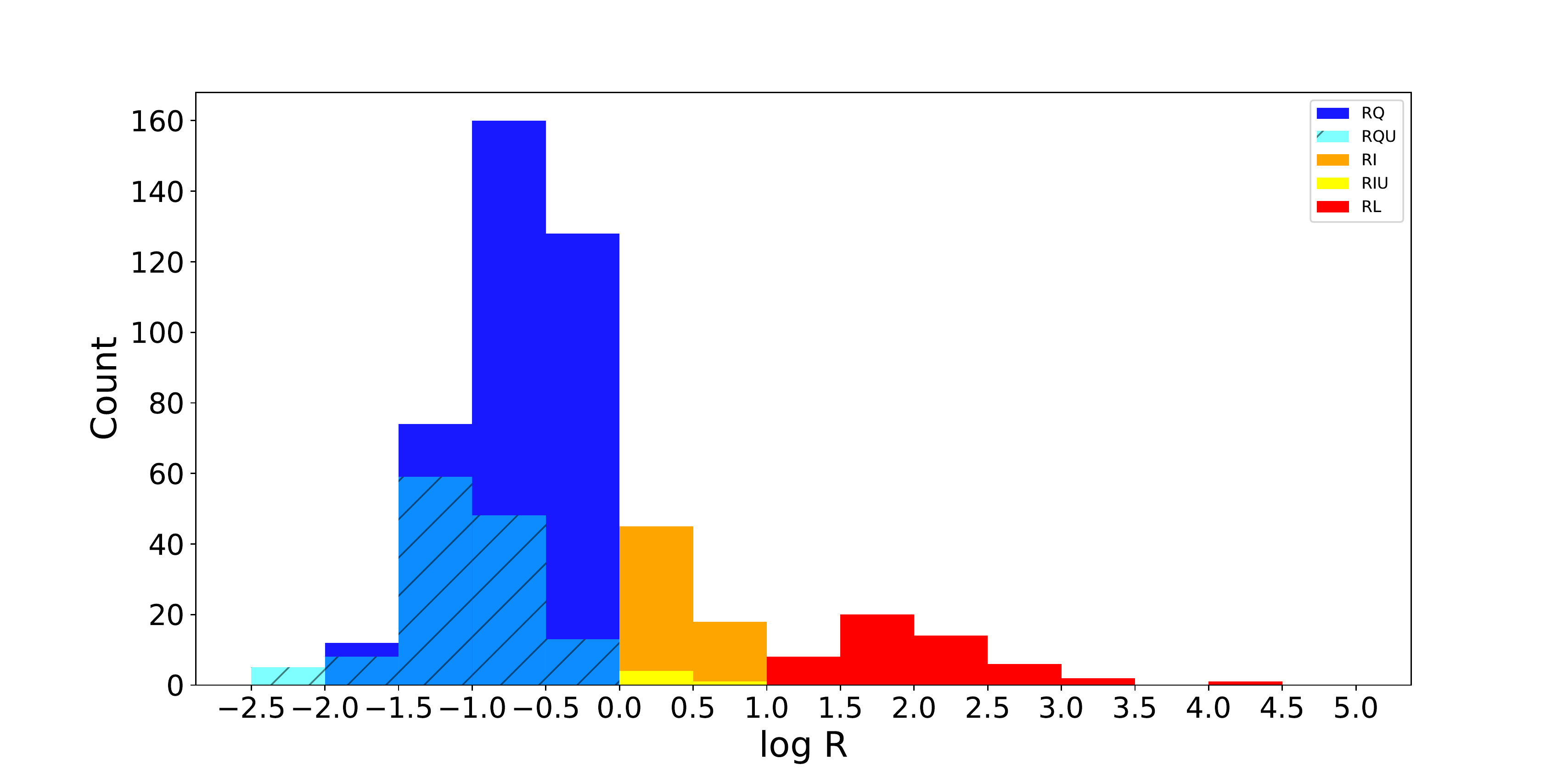}
\caption{The distribution of the radio loudness, $R = F_{1.4}/F_{\nu_{\rm W3}}$, for our sample.
  Besides showing objects with radio detections and categorized as: $R< 1$ as radio-quiet (RQ); $1 < R < 10$ as radio-intermediate (RI); and $R > 10$ as radio-loud (RL), we also present sources named as radio undetected (RQU and RIU).}

\label{img_hist_r1}
\end{figure*}

\section{THE DATA AND SAMPLES}

\subsection{X-ray data}
\label{subsec:Xray_data}

The {\it Swift}/Burst Alert Telescope (BAT) has carried out an all-sky survey
in the hard X-ray range (14--195 keV) and in the first 70 months of observation, it 
catalogued 1210 objects \citep{2013ApJS..207...19B}. The primary sample of AGNs
used in our research is derived from the BAT AGN Spectroscopic Survey (BASS),
from the studies conducted by \cite{2017ApJS..233...17R} where authors presented
X-ray spectral parameters for 838 AGNs.

\cite{2017ApJS..233...17R} also used  data from {\it Swift}/XRT, {\it XMM- Newton},
ASCA, {\it Chandra}, and Suzaku observations at energies $\le 10$ keV in order to
derive the column density and reflection features of {\it Swift}/BAT AGNs.
Using their data on column density we reduced the sample to include only
the Compton-thin AGNs ($\log N_{H} < 24$) therefore excluding objects with
significantly absorbed hard X-rays. This criterion restricted our sample to 776 sources. 

Our next step was to reject objects classified as blazars, since X-rays  
from these objects are likely to be dominated by the Doppler-boosted 
radiation from relativistic jets pointing towards us rather than by accretion flows.
The blazars were selected by using the \cite{2017ApJS..233...17R} catalogue,
where they were identified by cross-matching of {\it Swift}/BAT catalogue with 
the Roma-BZCAT 26 catalogue of blazars \citep{2015ApSS35775M}.  
Rejection of blazars left us with 664 objects.

\subsection{Radio data}
\label{subsec:radio_data}

Radio data for our sample, which includes sources from the Northern as
well as Southern hemisphere, were collected from two radio catalogues:
National Radio Astronomy Observatory (NRAO) Very Large Array (VLA) Sky
Survey (NVSS;  \citealt{1998AJ....115.1693C}) and Sydney University Molonglo Sky
Survey (SUMSS; \citealt{1999AJ....117.1578B,2003MNRAS.342.1117M}).
NVSS has been carried out at 1.4 GHz while SUMSS data is given at 843 MHz.
Both these surveys have similar angular resolution (45 arcsec FWHM for NVSS
and $45 \times 45 {\rm cosec} |\delta|$ arcsec$^2$ for SUMSS) and sensitivities ($\sim 2.5$ mJy).   

Since AGNs can have multiple radio sources associated with them, we ensured
to account for various types of morphologies (single or multiple, e.g. double
and triple, matches). Below we describe our method in more detail.  

As the first step in finding the NVSS associations for the objects in our
sample we decided to use a matching radius of 3 arcmin from the optical
coordinates of each source. Objects which had only a single association
within 3 arcmin were further matched with a closer radius of 30 arcsec
and if the radio association was still found within this radius the source
was classified as NVSS single. For objects with multiple matches within 3
arcmin we extracted maps of 0.45 deg\,$\times$\,0.45 deg in size from the
NVSS Postage Stamp Service\footnote{https://www.cv.nrao.edu/nvss/postage.shtml}
and visually inspected whether associations found by us actually do or do not belong to the
source. In order to do so we used the NRAO AIPS (Astronomical Image Processing System)
package and NED (NASA/IPAC Extragalactic Database)\footnote{https://ned.ipac.caltech.edu/}
which allowed us to distinguish incorrect matches. Additionally, when the radio morphology
was extended beyond the matching radius of 3 arcmin, the radius was gradually increased by
1 arcmin as long as the associations for the whole structure were found. The flux associated
with all the matches contributing to the given source was totalled up and assigned to each
object in our sample.

Since SUMSS has a similar resolution to NVSS, searching for radio associations in this catalogue
followed the same procedure as adopted for NVSS catalogue with radio maps extracted from the SUMSS
Postage Stamp Download\footnote{http://www.astrop.physics.usyd.edu.au/cgi-bin/postage.pl}.
Data given in this survey provides fluxes at 843 MHz which were re-calibrated to 1.4 GHz
using $\alpha_{\rm radio} = 0.8$ (with the convention $F_{\nu} \propto \nu^{-\alpha}$ , e.g. \citealt{1979A&A....80...13B,2008AJ....136..684K}).

\begin{figure*}
\centering
\includegraphics[scale=0.66]{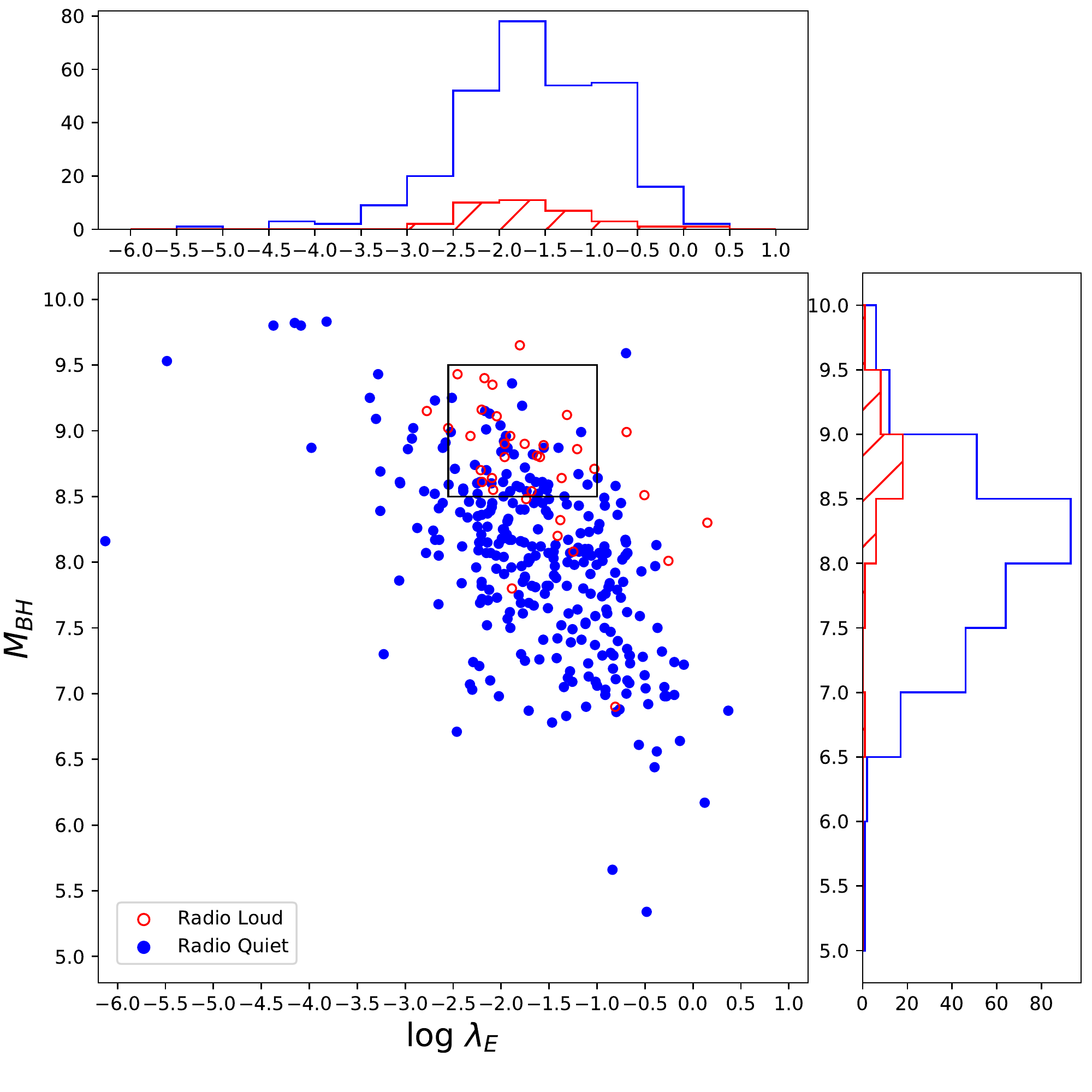}
\caption{The distribution of the black hole masses, $M_{\rm BH}$, and Eddington ratios, $\lambda_{\rm E}$, for RL (empty red) and RQ (filled blue) objects in our sample. The inlaid box represents the ranges of $M_{\rm BH}$ and $\lambda_{\rm E}$ cut-offs we focus on in this paper.}
\label{img_hist_mbh_lam}
\end{figure*}

\subsection{Mid-infrared data}
\label{subsec:MIR_data}

In order to obtain the infrared data for the sources in our sample we used the AllWISE Data
Release \citep{2013yCat.2328....0C}. The AllWISE release builds upon the work of the Wide-field
Infrared Survey Explorer mission ({\it WISE}; \citealt{2010AJ....140.1868W}) which observed the sky
at 3.4, 4.6, 12 and 22 ${\upmu}$m ({\it W1, W2, W3 and W4} band, respectively) and combines it with
the data from the WISE cryogenic and NEOWISE \citep{2011ApJ...731...53M} post-cryogenic survey
phases giving the best view of the mid-infrared (MIR) sky available till date. For determining
the counterparts for our sources we decided on 5 arcsec as the matching radius which arises
from the fact that the angular resolution of WISE is 6.1, 6.4, 6.5 and 12.0 arcsec in the
four bands, respectively. Using a larger matching radius would result in duplicate WISE
sources around our AGNs and higher rate of false positives in associations.

In order to accomplish our goal to use MIR data to determine 
radio loudness and bolometric luminosities regardless the optical type of AGNs
we decided to use only data from the {\it W3} band. That is because at shorter 
wavelengths (those covered by {\it W1} and {\it W2} bands) the dusty, circumnuclear
tori are optically thick and are radiating anisotropically
\citep{2011ApJ...736...26H,2015ARA&A..53..365N}, while 
measurements in {\it W4} band are affected by much larger errors than in {\it W3} band.
The monochromatic flux, $F_{\nu_{\rm W3}}$, is calculated using relation
$\log F_{\nu_{\rm W3}} = -(53.774 + m_{\rm W3})/2.5$, where  $m_{\rm W3}$
is the magnitude provided in the WISE catalogue.

\subsection{Constructing the RL and RQ samples}
\label{subsec:constructing_RL-RQ}

The radio loudness parameter is defined by us as  
$R = F_{1.4}/F_{\nu_{\rm W3}}$,
where $F_{1.4}$ and $F_{\nu_{\rm W3}}$ are the monochromatic fluxes at 1.4 GHz and 
$2.5 \times 10^{13}$ Hz, respectively. This is related to the definition given
by \cite{1989AJ.....98.1195K} as $R_{\rm KL} = F_5/ F_{\nu_{\rm B}}$,
where $F_5$ and $F_{\nu_{\rm B}}$ are the monochromatic fluxes at 5 GHz and 
$6.8 \times 10^{14}$ Hz, respectively. Both these quantities are
linked by assuming radio spectral index $\alpha_{\rm 1.4 -5} = 0.8$
and $\alpha_{\rm \nu_B-\nu_{W3}} = 1$ \citep{2006ApJS..166..470R}, and with them 
the relation between our radio loudness $R$ and the Kellerman parameter $R_{\rm KL}$
is found as $R \approx 0.1 \times R_{\rm KL}$.

Using the division corresponding  to that based on the \cite{1989AJ.....98.1195K} 
definition of radio loudness we define the following subsets in our sample: $R< 1$
as RQ; $1 < R < 10$ as radio-intermediate (RI); and $R > 10$ as RL. The distribution of the radio loudness of the sources in our sample is shown in
Fig \ref{img_hist_r1}. Together with RQ, RI and RL subsamples we present RQ and
RI sources for which the radio associations were not found. They are named as
RQU and RIU objects (with the radio flux as 2.5 mJy), respectively. 

\begin{table*}
\begin{center}
\caption{Steps involved in construction of our sample.}
\label{tbl_selection}
\begin{tabular}{c c } 
 \hline
Number of objects & Method
 \\ 
 \hline\hline
$838$ & Sample from \cite{2017ApJS..233...17R}\\
$776$ & Cut-off of $\log N_H < 24$ \\  
$664$ & Removing blazars \\ 
$630$ & Valid w3 magnitude \\
$560$ & Restricting our study to RL and RQ sources \\
$328$ & Valid $M_{\rm BH}$ estimation \\ 
$70$ ($24$ RL, $46$ RQ) & Cut-off of $8.48 \le \log M_{\rm BH} \le 9.5$  and $-2.55 \le \log \lambda_{\rm E} \le -1$\\
 \hline
\end{tabular}
\end{center}
\end{table*}

We hereafter reject from our further analysis RI AGNs and those radio-undetected
objects which have radio loudness  upper limits $> 1$. This assures
of having two `clean' subsamples, those with $R>10$ where radio emission is 
expected to be dominated by jets, and those
with $R<1$, where the total radio emission flux is likely to be dominated by 
star-forming regions  (e.g. \citealt{2011ApJ...739L..29K}), accretion disc winds
(e.g. \citealt{2016MNRAS.459.3144Z}), and/or accretion disc coronae (e.g. \citealt{2016MNRAS.459.2082R}). The resulting source count is 51 RL AGNs and 509 RQ AGNs.

\section{\textbf{\emph{Swift}}/BAT AGNs IN ${M_{\rm BH} - {\lambda}_{\rm E}}$ PLANE}
\subsection{The BH masses and Eddington ratios}

Black hole masses for the objects in our sample were taken from \cite{2017ApJ...850...74K}.
For AGN type 1 they were obtained using broad emission lines H$\beta$, 
for AGN type 2 -- using stellar velocity dispersion. For some AGNs in our sample
lacking information on BH masses in \cite{2017ApJ...850...74K} we took them from
\cite{2015MNRAS.447.1289P}. Together we gathered BH mass data for 328 AGNs
(293 for RQ and 35 for RL).

\cite{2017ApJ...850...74K} derived also bolometric luminosities. They 
assumed that bolometric luminosity is traced by the hard X-ray luminosities
with respective  bolometric correction. Noting however, that previous
studies indicated that RL AGNs are X-ray-louder than RQ AGNs,
we decided to use as a tracer of bolometric luminosity the MIR luminosities
(see e.g. \citealt{2012MNRAS.426.2677R,2015MNRAS.447.1184F}). This 
choice can be justified by noting: (i) that the MIR emission in AGNs is dominated by dust 
heated by central engine (e.g. \citealt{2015MNRAS.449.1422M,2017Natur.549..488R,2017ApJ...838L..20H,2018ApJ...862..118Z});
(ii) that the  'effective' covering factors of dusty ``obscurers'' (circumnuclear tori/winds),
defined to be the fraction of the optical/UV radiation converted by dust
to the IR radiation, are very similar in RL and RQ AGNs 
\citep{2013MNRAS.430.3445M,2016MNRAS.461.2346G, 2017ApJ...849...53H}
and (iii) that at $\lambda > 10\mu $m the AGN dusty 
obscurers are optically thin and  radiate isotropically (e.g. \citealt{2015MNRAS.449.1422M}).

The bolometric luminosity calculated by us is $L_{\rm bol} = K_{\rm W3} \nu_{\rm W3} L_{\nu_{\rm W3}}$, where
$L_{\nu_{\rm W3}} = 4\pi d_{L}^{2}F_{\nu_{\rm W3}}/(1+z)^{1-\alpha_{\rm IR}}$ {and
$K_{\rm W3}= 8.4$, where $\lambda_{W3} = 12\upmu$m \citep{2006ApJS..166..470R}.
Since dependence of this formula on $\alpha_{\rm IR}$ for small redshifts is 
very weak, we adopted $\alpha_{\rm IR} = 1$ \citep{2015MNRAS.449.1422M}.}
And combining this with
BH masses and definition of Eddington luminosity we obtain Eddington ratios,
$\lambda_E = L_{\rm bol}/L_{\rm Edd}$.

{Distribution of AGNs in the plane '$M_{\rm BH}$ -- $\lambda_E$'
is shown in Fig. \ref{img_hist_mbh_lam}.  As we can see, there is a deficiency of AGNs in the upper-right corner of the figure  
and in the lower-left corner of the figure. The low representation
of AGNs at large Eddington ratios and large  BH masses can be explained
as resulting from the steep mass function of AGN black holes (see
review by \citealt{2012AdAst2012E...7K}) and downsizing effect \citep{2012MNRAS.419.2797F},
while poor representation of AGNs at low Eddington 
ratios and small BH masses is associated  with the flux limit of {\it Swift}/BAT 
survey.  Furthermore, we can see very poor representation of RL AGNs with BH masses 
smaller than $10^8 M_{\odot}$ at any Eddington ratio. Such a dependence of 
radio loudness on BH mass, albeit well documented ( \citealt{2004MNRAS.353L..45M,2017ApJ...846...42K} and refs. therein), is still lacking consensus
regarding its explanation \citep[see e.g.][]{2011MNRAS.416..917C,2013ApJ...764L..24S,2016ApJ...833...30C}.
The objects at $\lambda_{\rm E} < 10^{-3}$ should not
be considered here, since the method applied by us for deriving bolometric luminosity
is expected to work only in the regime of radiatively efficient accretion flows.

\subsection{Luminous RGs and their 'close' RQ counterparts -- the GSRM sample}

Hereafter we focus on the RL and RQ AGNs covering similar ranges of BH masses 
and Eddington ratios and restrict our studies to AGNs with
$8.5 \le \log M_{\rm BH} \le 9.5$ and  $-2.55 \le \log \lambda_{\rm E} \le -1$. This
subset of AGNs is marked in Fig. \ref{img_hist_mbh_lam} by an inlaid box 
and is named by us the GSRM sample. It includes 70 objects consisting of 24 RL 
and 46 RQ AGNs, and detailed data on those objects are provided in Table 
\ref{tbl_sample}. Their redshift distributions are shown in Fig. \ref{img_z_dist}.
We see a significant overlap between them, and their median values
are 0.084 for RL AGNs and 0.070 for RQ AGNs. 

Distributions of radio loudness, $R$, of the GSRM RQ and RL subsamples 
are presented in Fig. \ref{img_hist_Rloud}.
The median value of the radio loudness of  RL bjects is 122 and of 
RQ objects is by a factor $\sim 500$ smaller. 

\begin{figure}
\centering
\includegraphics[scale=0.4]{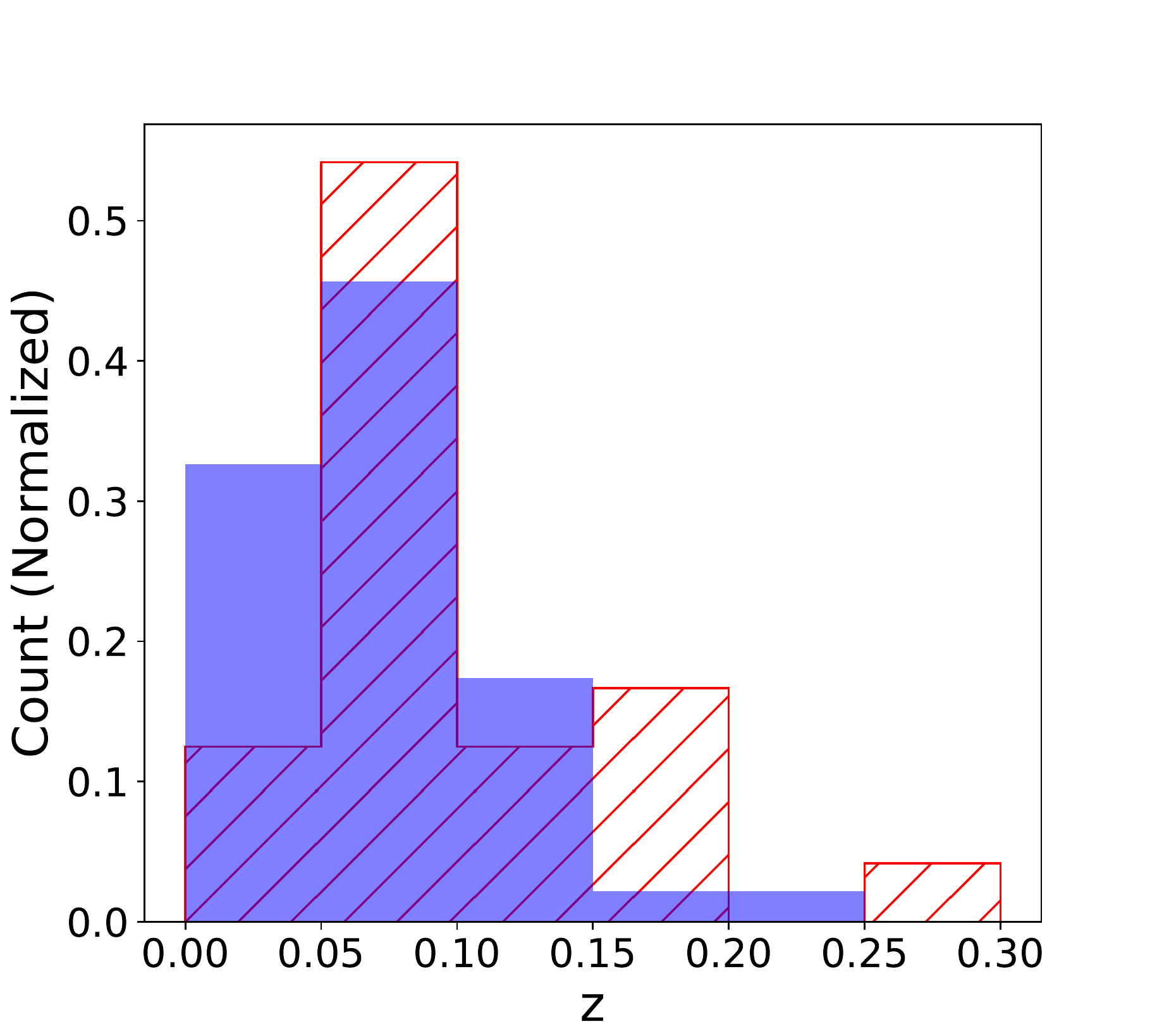}
\caption{The normalized distribution of the redshift of RL (hatched red) and RQ (filled blue) samples of GSRM AGNs.}
\label{img_z_dist}
\end{figure}

For each of the sources in our sample, we extracted radio maps from NVSS and
  SUMSS surveys (as described in section \ref{subsec:radio_data}).
  We found that most of RL AGNs in the GSRM sample have FR\,II radio morphology.
  20 of radio galaxies in the GSRM sample were also identified with  {\it Swift}/BAT AGNs by \cite{2016MNRAS.461.3165B}. 
Among them we can distinguish three FR\,II objects with peculiar radio structure: SWIFT J2359.3-6058 being a giant radio galaxy (GRG, \citealt{2015MNRAS.449..955M}); SWIFT J2223.9-0207
  which is a well-aligned double-double giant radio galaxy (DDRG, \citealt{2006MNRAS.366.1391S});
  and one X-shaped source, SWIFT J1952.4+0237 \citep{2002A&A...394...39C}.
  Four GSRM RL AGNs are not identified by \citeauthor{2016MNRAS.461.3165B} They are: SWIFT J0021.2-1909,
  which is a triple in NVSS; SWIFT J0235.3-2934 with two radio matches in NVSS and extended (but not FR\,II)
  structure; and SWIFT J1207.5+3355 and SWIFT J2351.9-0109 -- both found as singles on NVSS map and
  probably FR\,II on FIRST (Faint Images of the Radio Sky at Twenty-centimeters; \citealt{1995ApJ...450..559B}) maps. Detailed studies of radio sources associated with GSRM AGNs will
  be investigated in our future paper, where the issue of radio bimodality will be addressed.

It is worth mentioning here that the number ratio of
RL to RQ objects in the GSRM sample is $\sim 1:2$. This ratio is much 
  larger than the ratio $\sim 1:10$ of RL quasars to RQ quasars \citep{2006AJ....131..666D}.
This  difference can be explained by noting that in our
sample all AGNs host very massive BHs and that the fraction of RL AGNs among
AGNs with $M_{\rm BH} > 10^8 M_{\odot}$ is larger than among AGNs with smaller 
BH masses (see e.g. \citealt{2004MNRAS.353L..45M}). However, one cannot exclude a possibility that the fraction of the RL AGNs depends also on 
the Eddington ratio being larger at lower accretion rates (see \citealt{2007ApJ...658..815S,2017MNRAS.466.2294R} and references therein).

\begin{figure}
\centering
\includegraphics[scale=0.35]{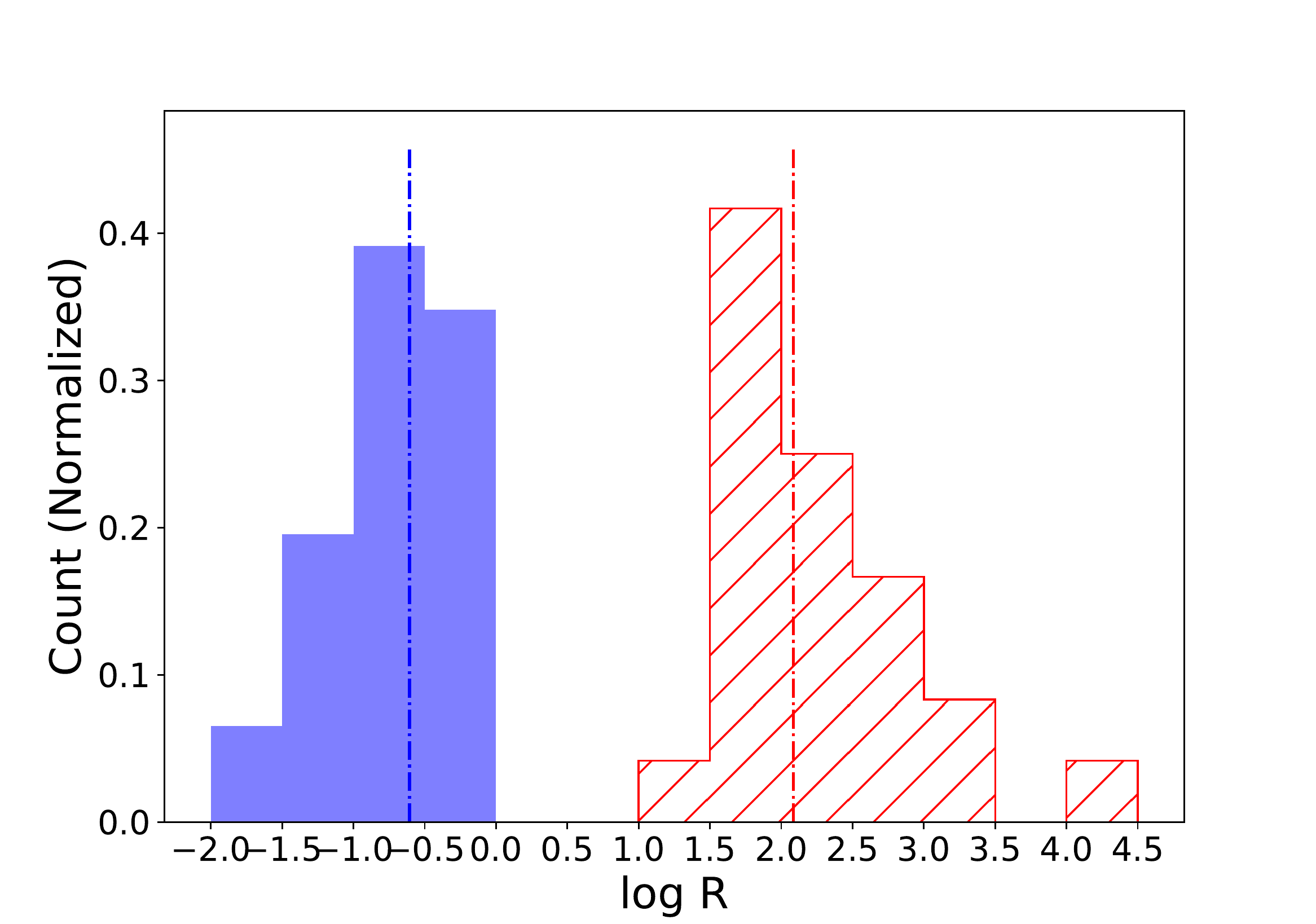}
\caption{The normalized distribution of RL (hatched red) and RQ (filled blue) subsamples of GSRM AGNs in radio loudness. The median values are presented as dashed lines for both samples.}
\label{img_hist_Rloud}
\end{figure}

\begin{figure}
\centering
\includegraphics[scale=0.4]{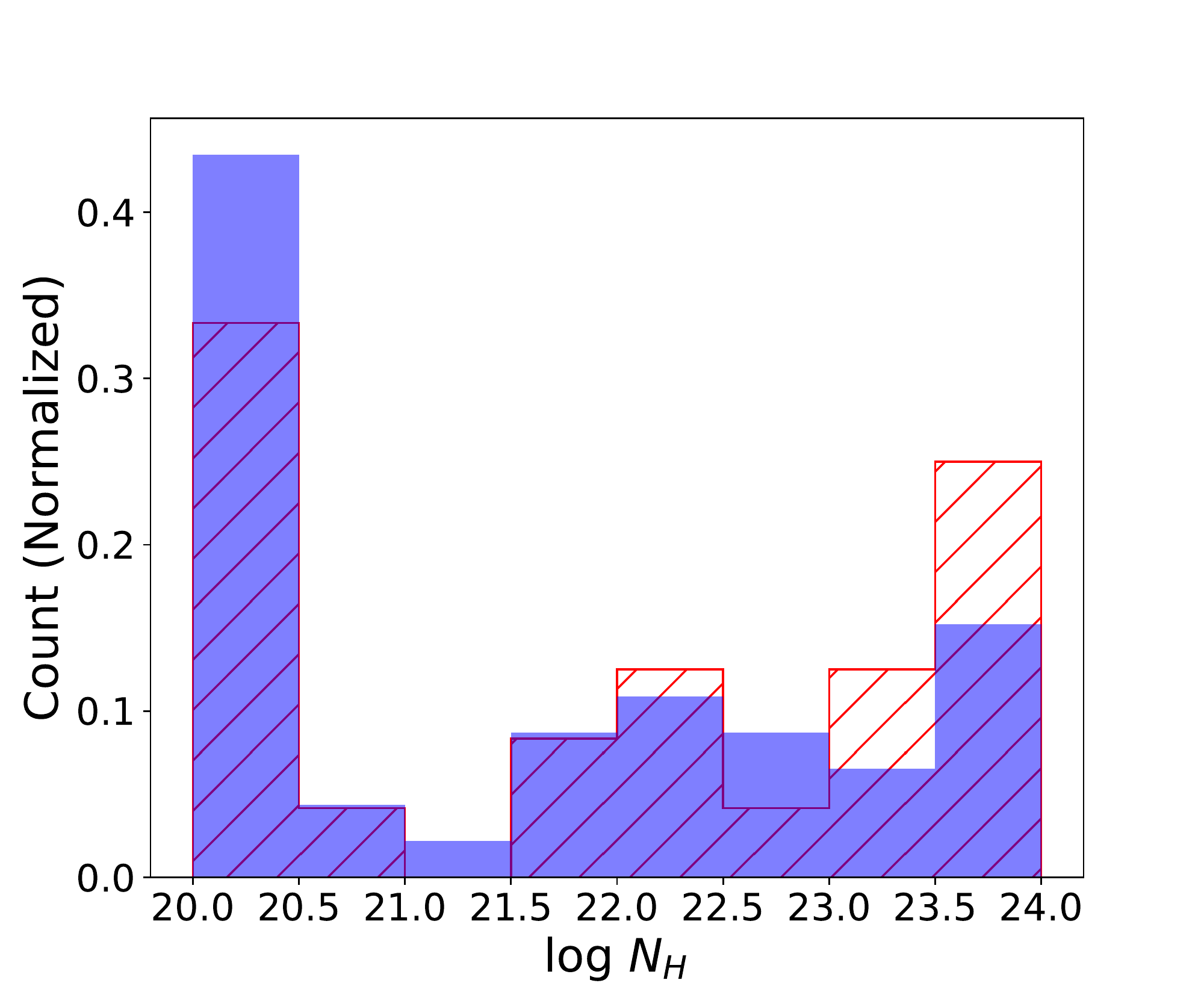}
\caption{The normalized distribution of the column density, $N_H$, of RL (hatched red) and RQ (filled blue) samples of GSRM AGNs.}
\label{img_nh_dist}
\end{figure}


Fig. \ref{img_nh_dist} shows the distribution of column density, $N_H$, for the
RL and RQ samples. We performed the Kolmogorov-Smirnov test to the distributions 
of column density for the RL and RQ samples and observe that they are very similar
(see Table \ref{tbl_ks}). This indicates  similar covering factors 
of circumnuclear dusty obscurers and validates our method of using MIR to determine
the bolometric luminosity with the same bolometric correction 
for RL and RQ AGNs (see also previous subsection). However it should be noted 
that at $\log N_H < 20.5$ the column densities are dominated by the Milky Way and
host galaxies rather than by the AGN circumnuclear obscurers.

\section{COMPARISON OF X-RAY PROPERTIES}

\subsection{The X-ray loudness}
\label{subsec:X-ray_loudness}

The definition of X-ray loudness for our sample is adopted as the ratio of hard X-ray luminosity to bolometric luminosity, $L_{14-195}/L_{\rm bol}$. 
Distributions of X-ray loudness
for RL and RQ AGNs are  presented in Fig.~\ref{img_hist_xloud}. As we can see 
the X-ray loudness distribution
of RQ AGNs is broader  and extends to much lower values than 
the distribution of RL AGNs. The median value of X-ray loudness is $\sim 0.4$
and $\sim 0.2$.
for RL and RQ sources, respectively. Therefore we observe that
the RL objects are on average about two times X-ray-louder than their 
RQ counterparts in the 14 -- 195 keV band. The low p-value of 0.0007 of the
  K-S test indicates that the X-ray loudness distribution for the RL and RQ samples
  are quite different (see Table \ref{tbl_ks}).
This difference could be larger by considering the possibility of greater 
incompleteness of the X-ray-detected RQ AGNs than of RL AGNs. However
location of members of the GSRM sample in the $L_{14-195}$--redshift plane 
relative the flux-limited luminosity curve (see Fig. \ref{img_KcorXray_redshift})
indicates that within the ranges of BH masses and Eddington ratios 
covered by this sample this incompleteness is rather negligible.

\begin{figure}
\centering
\includegraphics[scale=0.4]{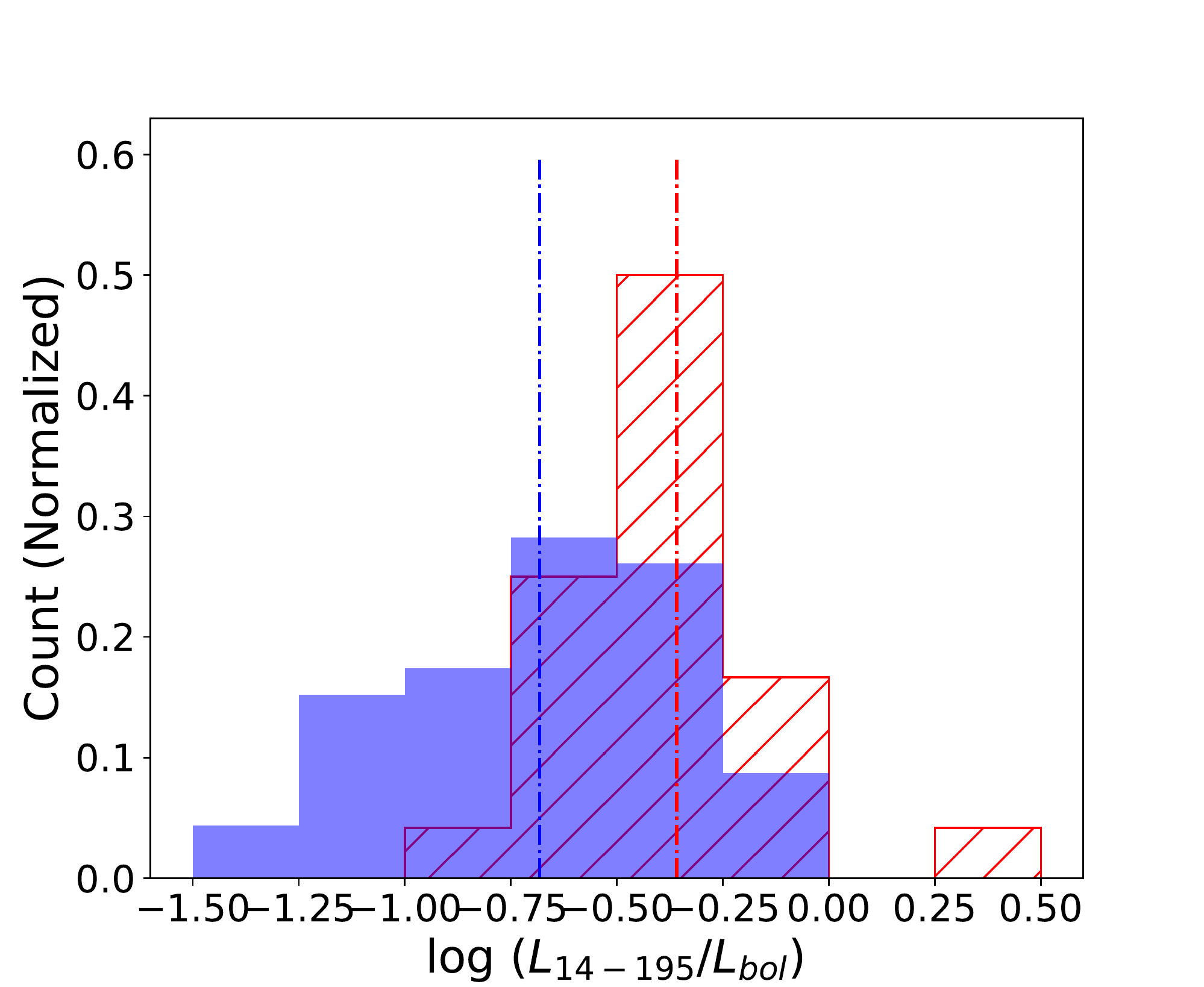}
\caption{The normalized distribution of RL (hatched red) and RQ (filled blue)
  subsamples of GSRM AGNs in X-ray loudness which is defined as the ratio of hard X-ray
  luminosity to bolometric luminosity.
  The median values are presented as dashed lines for both samples.}
\label{img_hist_xloud}
\end{figure}

\begin{figure}
\centering
\includegraphics[scale=0.4]{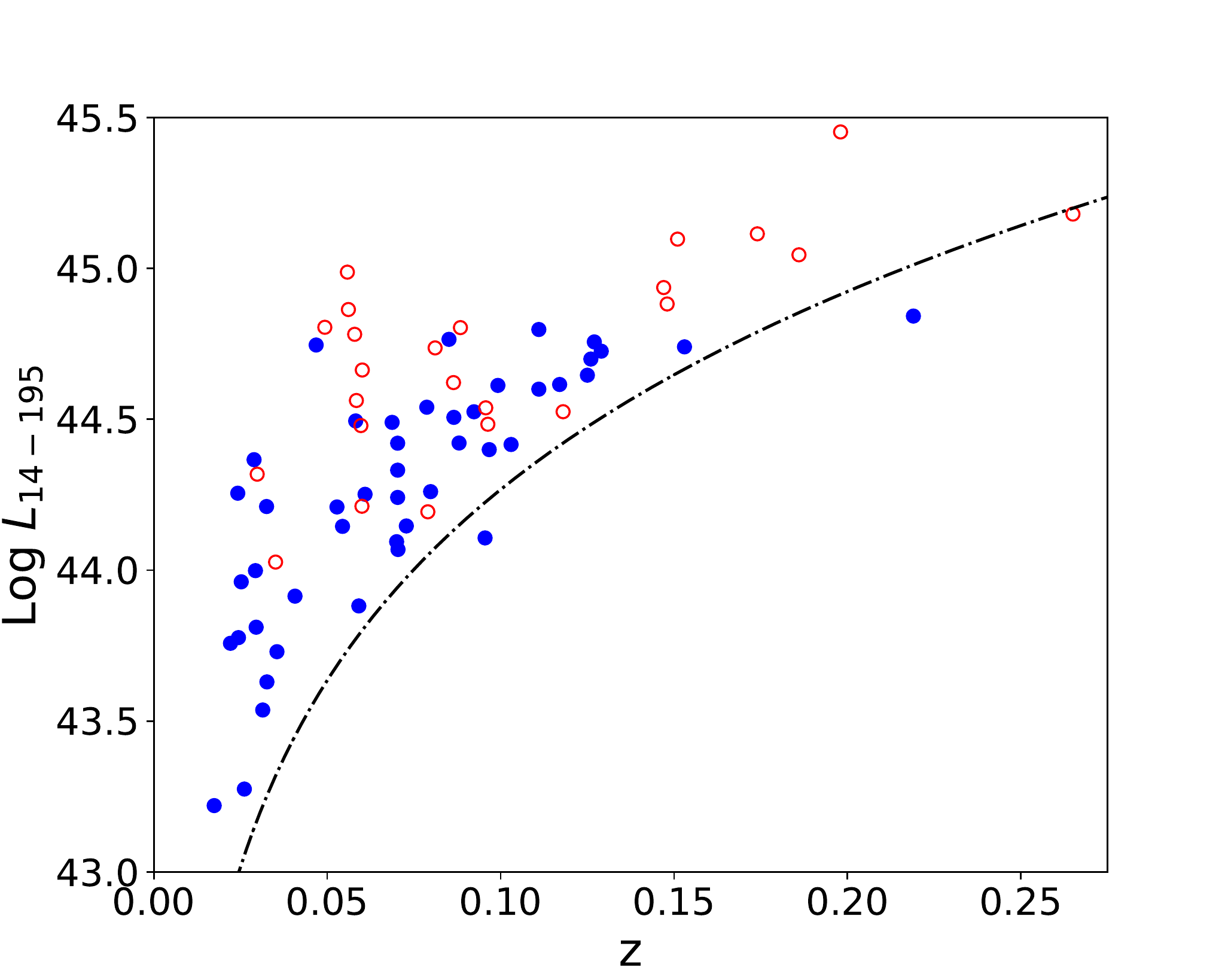}
\caption{The distribution of hard X-ray luminosity versus redshift of RL (empty red) and RQ (filled blue)
  subsamples of GSRM AGNs.}
\label{img_KcorXray_redshift}
\end{figure}

\subsection{The spectral slopes}
\label{subsec:spectral_slopes}

In our study we decided on using the photon index $\Gamma_{14-195}$ values provided
by \cite{2017ApJS..233...17R} instead of those from BAT sample \citep{2013ApJS..207...19B}
where the obscuration by dusty torus was not taken into account. The distribution of $\Gamma_{14-195}$
for the RL and RQ sources is shown in Fig. \ref{img_hist_gamma}.
We clearly see a significant overlap between both samples with very similar
median values being 1.740 and 1.735 
for RL and RQ sources respectively and the high p-value of 0.969 of the
K-S test indicates that the spectral index distribution for the RL and RQ samples is very similar (see Table \ref{tbl_ks}).

\begin{figure}
\centering
\includegraphics[scale=0.35]{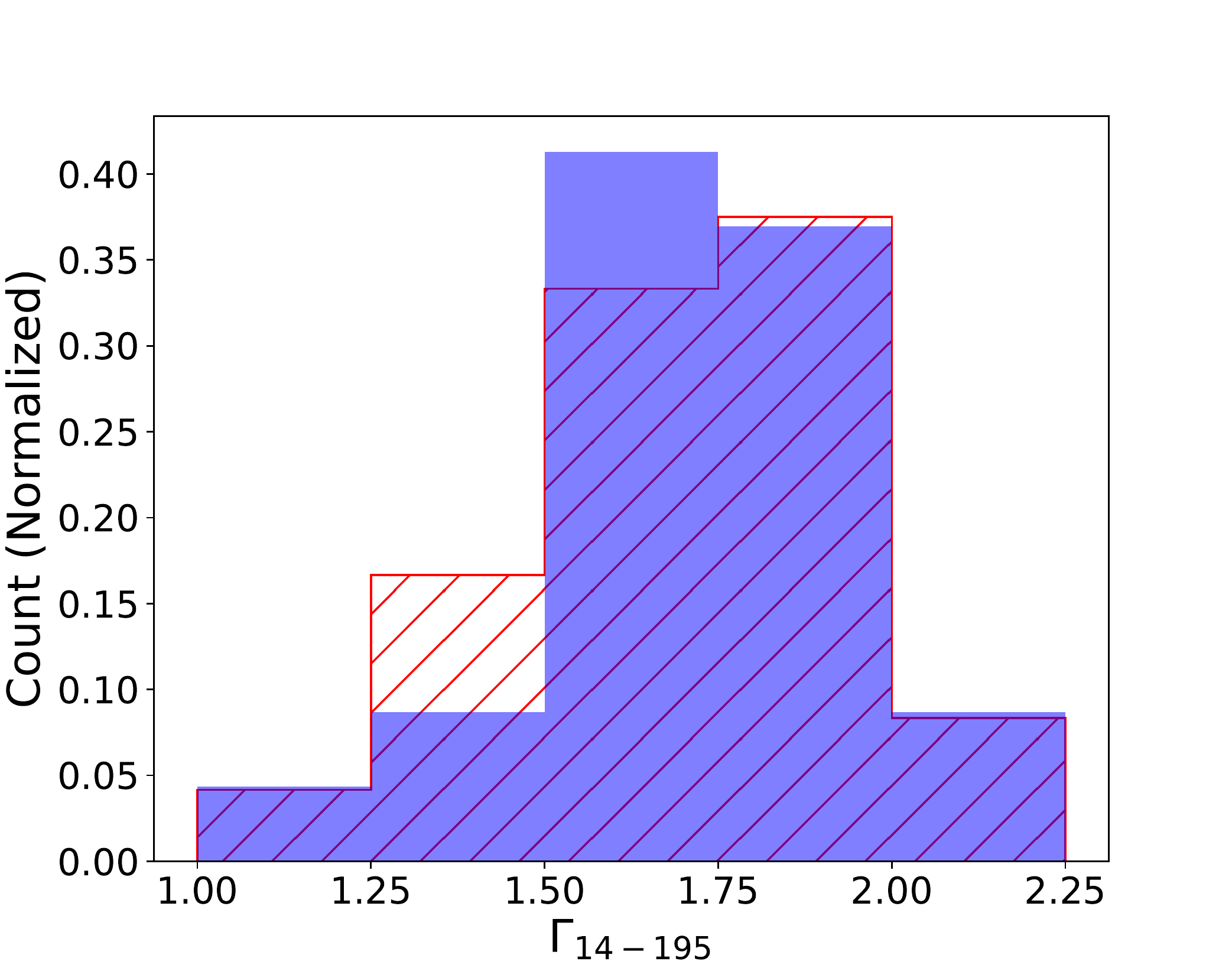}
\caption{The normalized distribution of the photon index, $\Gamma_{14-195}$, of RL (hatched red) and RQ (filled blue) subsamples of GSRM AGNs.}
\label{img_hist_gamma}
\end{figure}

\begin{table}
\begin{center}
\caption{Results of the Kolmogorov-Smirnov test performed on the distributions of column density, X-ray loudness, and spectral slope for the RL and RQ samples}
\label{tbl_ks}
\begin{tabular}{c c c} 
 \hline
Distribution & $D$  & $p-\rm value$ \\
\hline
\hline
$\log N_H$ & 0.190 & 0.571 \\
$\log L_{14-195}/L_{\rm bol} $ & 0.486 & 0.0007 \\
$\Gamma_{14-195}$ & 0.119 & 0.969 \\
\hline
\end{tabular}
\end{center}
\end{table}

Even though in many previous studies the emission of hard
X-rays was found to be higher for RL than RQ AGNs \citep[e.g.][and references therein]{1981ApJ...245..357Z,2011ApJ...740...29K,1987ApJ...313..596W,1998MNRAS.299..449W,2011ApJ...726...20M} 
until now none of 
them has focused on similar black hole mass and Eddington ratio ranges.
While considering this aspect one can notice no difference in
the photon index distribution between RL and RQ samples
which suggests a similar underlying mechanism in the production
of X-rays being independent on the radio loudness value.

\subsection{High-energy cut-offs and reflection features}
\label{subsec:high-energy-cutoff_reflection-features}

The high-energy cut-off values were, in similarity to the
photon gamma index, taken from \cite{2017ApJS..233...17R}.
Some of the sources have given only the lower limits while
for three RL and three RQ sources there are no data resulting in excluding
them in this analysis. The distributions of high-energy cut-off for
RL and RQ subsamples are presented in Fig. \ref{img_hist_ec_rl}
and \ref{img_hist_ec_rq}, respectively, and they are comparable.
Having the high-energy break for RL and RQ AGNs around the same place
implies that the mechanism that produces the X-rays is similar.

\begin{figure*}
\begin{subfigure}{0.49\linewidth}
\centering
\includegraphics[scale=0.25]{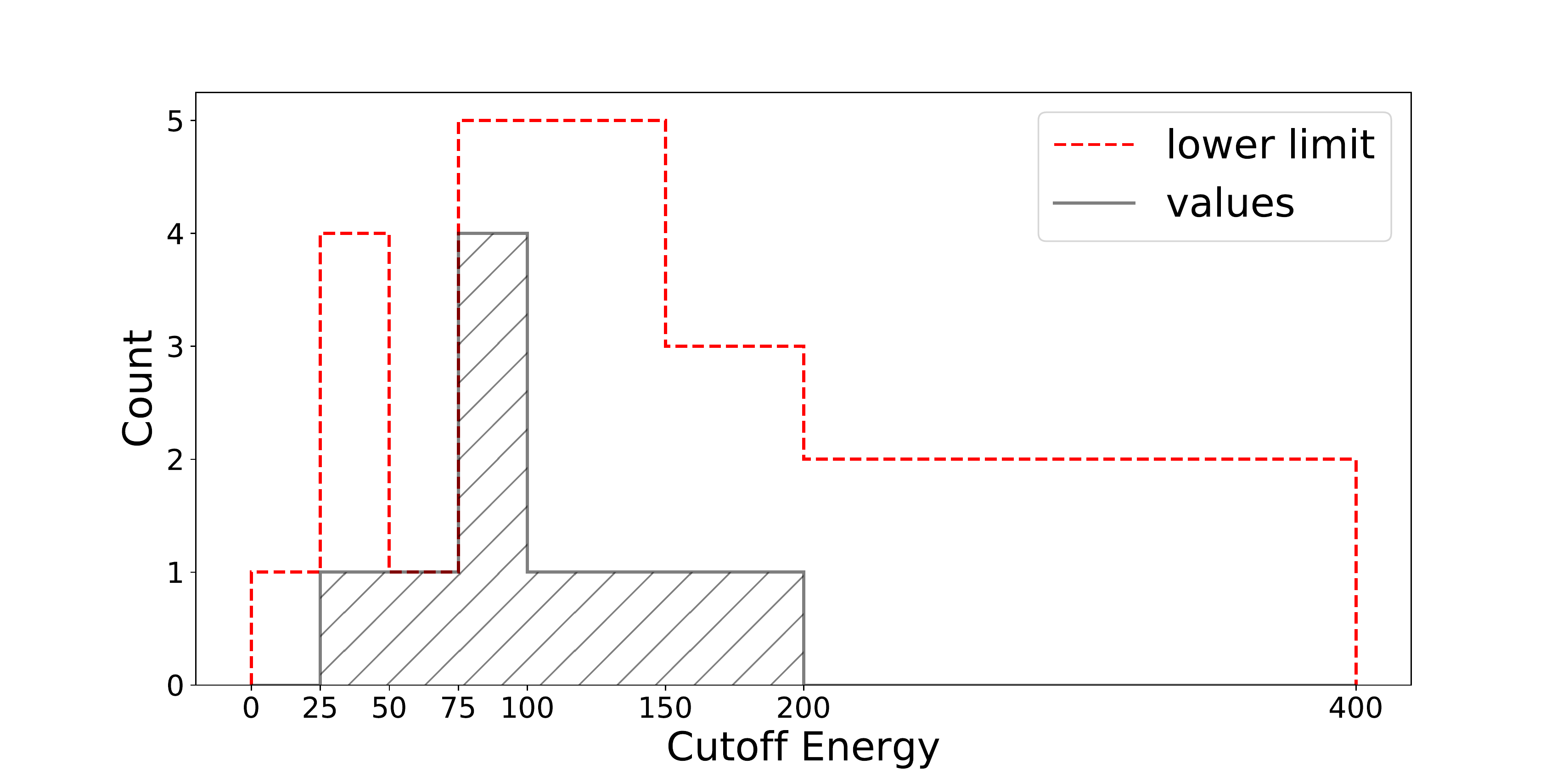}
\caption{}
\label{img_hist_ec_rl}
\end{subfigure}\hfill
\begin{subfigure}{0.49\linewidth}
\centering
\includegraphics[scale=0.25]{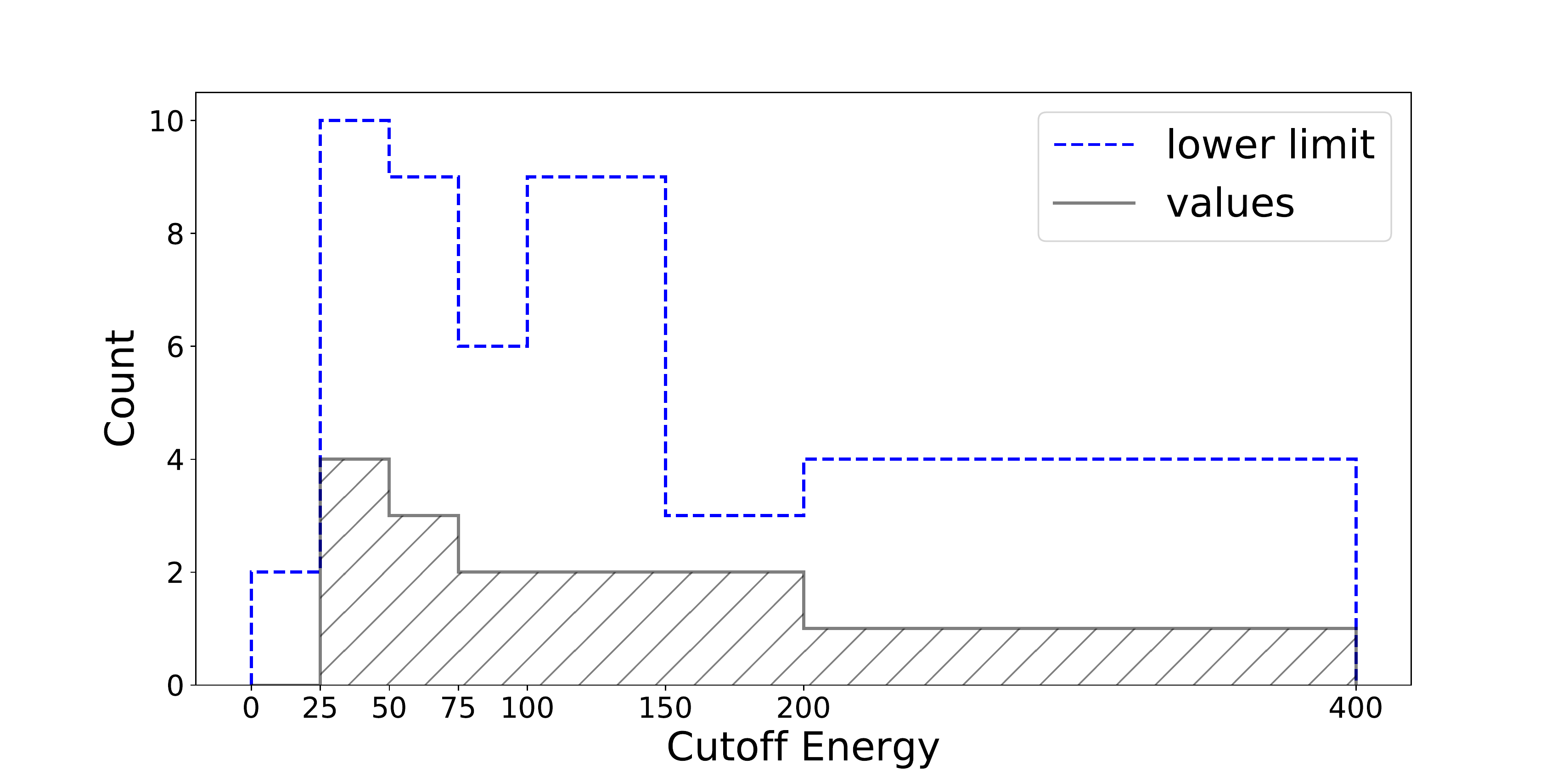}
\caption{}
\label{img_hist_ec_rq}
\end{subfigure}\hfill
\caption{Cut-off energy distribution. The plots show the values (black solid histogram) and the lower limits (dashed histogram). (a) Distribution of the cut-off energy for the RL AGNs of GSRM sample. (b) Distribution of the cut-off energy for the RQ AGNs of GSRM sample.}
\end{figure*}

Regarding the reflection features, we also use results 
obtained by \cite{2017ApJS..233...17R}. Those authors provide constraints on the 
reflection parameter $R_{\rm refl} = \Omega/2\pi$, where $\Omega$ is 
the solid angle subtended by the reflecting material,  calculated
using data from {\it Swift}/XRT and other X-ray telescopes covering 
the X-ray bands below 10 keV. Values of $R_{\rm refl}$ are found  for  
two RL AGNs and nine RQ AGNs, for others only upper limits are 
available (see  Fig. \ref{img_hist_rc_rl} and Fig. \ref{img_hist_rc_rq}).
The larger fraction of RQ AGNs with determined 
values of $R_{\rm refl}$ than of RL AGNs ($9/46$ in RQ AGNs versus $2/24$ in RL AGNs) 
and larger on average upper limits of $R_{\rm refl}$ in RQ AGNs  indicate 
smaller on average covering factor of reflecting material in RL AGNs. 
This is in agreement with results obtained by other studies of reflection 
features disregarding different selection methods of RL and RQ AGNs  \citep[e.g.][and
references therein]{1998MNRAS.299..449W,2011ApJ...726...20M}.
 
\begin{figure*}
\begin{subfigure}{0.49\linewidth}
\centering
\includegraphics[scale=0.25]{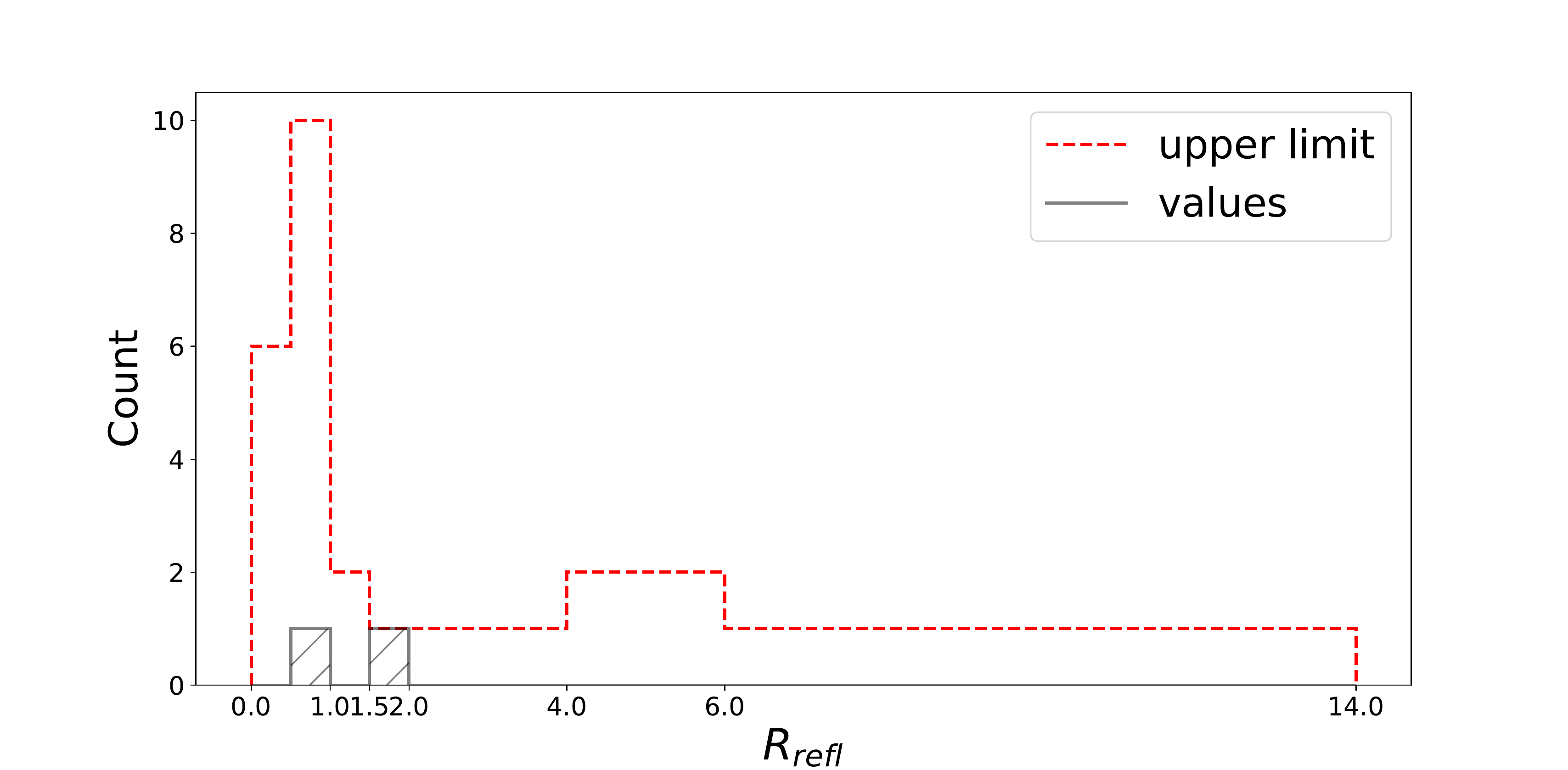}
\caption{}
\label{img_hist_rc_rl}
\end{subfigure}\hfill
\begin{subfigure}{0.49\linewidth}
\centering
\includegraphics[scale=0.25]{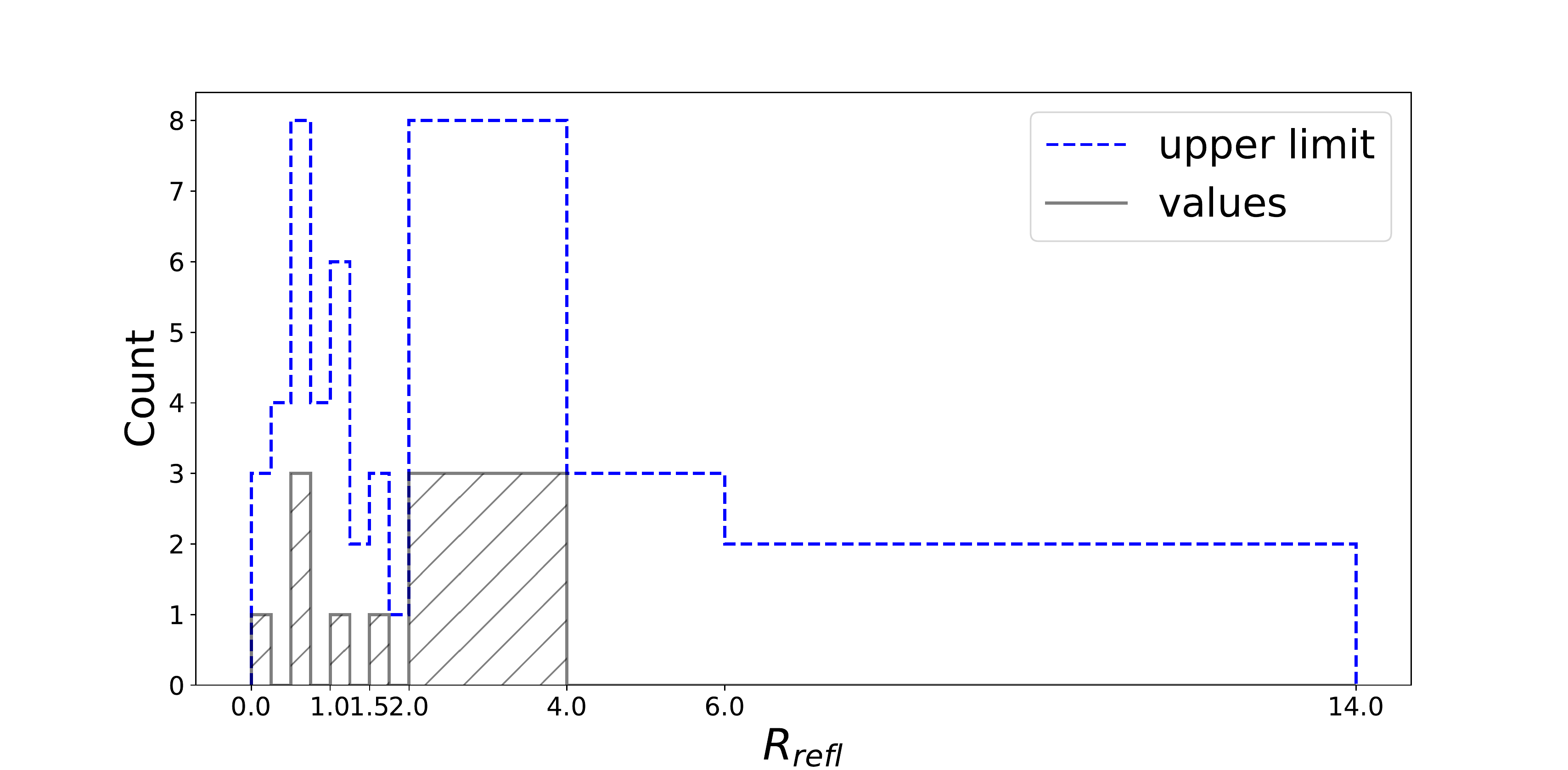}
  \caption{}
\label{img_hist_rc_rq}
\end{subfigure}\hfill
\caption{Reflection coefficient distribution. The plots show the values (black solid histogram) and the upper limits (dashed histogram). (a) Distribution of the reflection parameters for the RL AGNs of GSRM sample. (b) Distribution of the reflection parameters for the RQ AGNs of GSRM sample.}
\end{figure*}

\section{DISCUSSION}

Perhaps the largest puzzle in our understanding of active galaxies still remains
the very large diversity of their jet activity. This is particularly most apparent in AGNs with 
higher accretion rates, where power of a jet traced by the radio luminosity
can be compared with accretion power traced by optical luminosity of accretion 
discs. At each Eddington ratio the radio loudness spans at least by 
three to four orders of magnitude (e.g. \citealt{2007ApJ...658..815S}). 
Jets are observed also in some RI and RQ AGNs  (see e.g. \citealt{2016ApJ...832..163S} and references therein), but in many (probably most) of them the radio emission 
is likely to be dominated by star formation regions (SFR e.g. \citealt{2011ApJ...739L..29K}), the shocked accretion disc winds 
(e.g. \citealt{2016MNRAS.459.3144Z}), and/or accretion disc
coronae (e.g. \citealt{2016MNRAS.459.2082R}). Hence the range of radio 
luminosities associated with the jet activity is larger than the one 
determined  assuming that even in RQ AGNs radio emission is dominated by jets.
And regardless whether the jets are present and 
how powerful they are, quasar SEDs covering 
IR, optical and UV bands are very similar, and  
statistically significant differences appear only in X-ray bands 
\citep{1994ApJS...95....1E,2006ApJS..166..470R,2011ApJS..196....2S}. In particular 
the RL quasars are found to have on average: 
(1) larger X-ray-to-optical luminosity ratios; (2) harder X-ray spectra; 
(3) weaker reflection features. Similar differences have been claimed
for lower redshift type 1 AGNs, those represented by broad-line
radio galaxies and RQ type 1 Seyferts and having on average lower Eddington
ratios than quasars \citep{2011ApJ...726...20M,2011ApJ...740...29K}.
In most of these studies the compared RL and RQ samples
were selected independently, without considering the differences of their average 
black hole masses and Eddington ratios. Since X-ray properties may depend on 
these parameters, we decided to compare X-ray properties of RLAGNs with
their RQ counterparts selecting them  to cover a similar range of BH masses and accretion rates,
namely $8.5 \le \log M_{\rm BH} \le 9.5$ and  $-2.55 \le \log \lambda_{\rm E} \le -1$.

We found that indeed RQ counterparts of luminous radio galaxies
have on average lower X-ray luminosities, but only by a factor $\sim 2$.
On the other hand, similar distributions of hard X-ray spectral slopes and of
high-energy spectral cuto-ffs of RL and RQ GSRM subsamples which in 
radio loudness differ by a factor $\sim 500$ implies 
that production of even very powerful jets  does not affect significantly the 
properties of the primary X-ray sources. This strongly suggests similar
mechanism and location of hard X-ray production in RQ and RL AGNs and 
therefore appears to exclude any significant contribution of jets to the
X-ray luminosities even in RL AGNs. A scenario favoured by detailed observations and studies
of individual BLRGs (e.g. \citealt{2000ApJ...537..654E,2007ApJ...659..235G,2009ApJ...700.1473S,2010ApJ...721.1340T,2013ApJ...772...83L})
seems to have hard X-ray emission result from Comptonization of optical/UV radiation produced by 
cold, truncated accretion disc by high-energy electrons in 
hot, geometrically thick  ADAFs between the cold disc and 
the black hole (for more complex structures of thermal
Compton models of X-ray production in AGNs see e.g. \citealt{2018MNRAS.480.1247K} and references therein).
In such models the high-energy cut-offs of the X-ray spectra are determined
by temperature of electrons heated in ADAFs by hotter ions.

The GSRM  sample selected by us is planned to be used for more 
detailed comparisons of RL and RQ AGNs, including detailed radio data, narrow emission lines, optical
properties of their host galaxies, environments, and in case of AGN type 1,
their SEDs covering NIR, optical, UV and soft-X-ray bands. 

\section{SUMMARY}

By matching the {\it Swift}/BAT catalogue with the NVSS and SUMSS catalogues
we selected RL and RQ samples of AGNs having detection
in the 14-195 keV band. Then, estimating their bolometric luminosities
using mid-IR data from WISE and  taking BH masses  
from \cite{2017ApJ...850...74K} and for some RL AGNs not included in
the paper by \citeauthor{2017ApJ...850...74K}~ from \cite{2015MNRAS.447.1289P},
we selected two respective subsamples covering similar ranges of BH masses and Eddington ratios, $8.5 \le \log M_{\rm BH} \le 9.5$ and $-2.55 \le \log \lambda_{\rm E} \le -1$. They are represented by 24 luminous 
radio galaxies and by 46 RQ AGNs. Both subsamples have similar
X-ray spectral slopes. They have also similar maximal values of X-ray loudness, 
however with RQ AGNs having lower average X-ray loudness than radio galaxies. 
Above similarities suggest that the location and mechanism of hard X-ray production are the same in radio galaxies 
and  in their RQ counterparts. The mechanism most likely
involves Comptonization of  optical/UV radiation of truncated cold accretion 
disc by hot electrons in centrally developed ADAFs.
In this scenario the larger on average
X-ray loudness of RL AGNs than of RQ ones can be explained assuming that BHs 
in RL AGNs rotate faster. This is because in such a case efficiency  of 
conversion of accretion energy to radiation in hot central portions
of accretion flows  is expected to be 
larger.  Furthermore, in a case of larger BH spins in radio louder AGNs one 
may expect to have in such objects larger central 
concentration  of the hard X-ray emission, which in turn may explain 
lower reflection features in luminous radio galaxies than in their RQ counterparts.

  

\section*{Acknowledgements}

We acknowledge financial support by the Polish National Science Centre grants 2016/21/B/ST9/01620 and 2017/25/N/ST9/01953,
GM acknowledges the support of the United States Department of Energy Office of Science contract to SLAC no. DE-AE3-76SF00515.









\newpage

\clearpage
\newgeometry{top=1cm,right=5.5cm}

\begin{sidewaystable}
\scriptsize
\begin{tabular}{c c c c c c c c c c c c c c c c c} 
\hline

BAT NAME  &  $z$  &  $\log L_{14-195}$ & $F_{1.4}$    & $F_{\nu_{\rm W3}}$ &  $\log R$  & Radio class & Optical type &  $\log M_{\rm BH}/M_{\odot}$  &  $L_{\rm bol}$ &  $\lambda_{\rm E}$  &  $L_{14-195}/L_{\rm bol}$  & $\Gamma_{14-195}$    &  $E_{\rm cut-off}$  &  $\log N_{H}$  &  $R_{\rm refl}$  \\
 \\ 
 & & $\rm erg\,s^{-1}$ & $\rm erg\,s^{-1} cm^{-2} Hz{-1}$ & $\rm erg\,s^{-1} cm^{-2} Hz{-1}$ & & & &  & $\rm erg\,s^{-1}$ & & & & keV & $\rm cm^{-2}$ &\\
 \hline\hline
SWIFT J0021.2-1909  & 0.0957 & 44.6 & 1.20E-23 & 2.26E-25 & 1.73 &  RL  & 2 & 9.16 & 45.1 & -2.2 & 0.31 & 1.7 &  $86$  & 22.0 &  $\le 0.2$ \\
SWIFT J0042.9-2332  & 0.0221 & 43.8 & 4.22E-25 & 1.17E-24 & -0.44 &  RQ  & 1 & 8.6 & 44.5 & -2.25 & 0.2 & 1.76 &  $\ge 115$  & 23.5 &  $0.6$ \\
SWIFT J0109.0+1320  & 0.0597 & 44.5 & 1.37E-22 & 3.72E-25 & 2.57 &  RL  & 2 & 8.48 & 44.8 & -1.74 & 0.44 & 1.97 &  $\ge 136$  & 23.8 &  $0.5$ \\
SWIFT J0134.1-3625  & 0.0298 & 44.3 & 5.09E-23 & 9.93E-25 & 1.71 &  RL  & 2 & 8.64 & 44.6 & -2.1 & 0.49 & 1.07 &  $75$  & 24.0 &  $\le 0.1$ \\
SWIFT J0215.6-1301  & 0.148 & 44.9 & 4.68E-23 & 1.48E-25 & 2.5 &  RL  & 2 & 8.81 & 45.3 & -1.63 & 0.41 & 1.87 &  $\ge 48$  & 23.2 &  $\le 0.8$ \\
SWIFT J0235.3-2934  & 0.06 & 44.2 & 2.47E-24 & 1.97E-25 & 1.1 &  RL  & 1 & 9.02 & 44.6 & -2.55 & 0.45 & 1.68 &  $\ge 366$  & 20.0 &  $\le 1.3$ \\
SWIFT J0249.1+2627  & 0.0582 & 44.5 & 2.22E-25 & 6.87E-25 & -0.49 &  RQ  & 2 & 9.15 & 45.1 & -2.17 & 0.26 & 1.78 &  $\ge 130$  & 23.6 &  $\le 0.2$ \\
SWIFT J0255.2-0011  & 0.0289 & 44.4 & 1.54E-24 & 1.65E-24 & -0.03 &  RQ  & 2 & 9.25 & 44.8 & -2.51 & 0.34 & 1.69 &  $\ge 28$  & 23.8 &  $\le 0.2$ \\
SWIFT J0300.0-1048  & 0.0326 & 43.6 & 1.75E-25 & 3.98E-25 & -0.36 &  RQ  & 1 & 8.71 & 44.3 & -2.48 & 0.2 & 1.98 &  $\ge 206$  & 22.0 &  $\le 1.1$ \\
SWIFT J0407.4+0339  & 0.0884 & 44.8 & 5.68E-23 & 7.59E-26 & 2.87 &  RL  & 2 & 8.61 & 44.5 & -2.2 & 2.08 & 1.29 &  $56$  & 23.6 &  $\le 0.8$ \\
SWIFT J0418.3+3800  & 0.0493 & 44.8 & 1.53E-22 & 6.91E-25 & 2.35 &  RL  & 1 & 8.8 & 44.9 & -1.96 & 0.75 & 1.76 &  $\ge 144$  & 21.9 &  $\le 0.5$ \\
SWIFT J0429.6-2114  & 0.07 & 44.1 & 1.03E-25 & 4.32E-25 & -0.62 &  RQ  & 1 & 8.82 & 45.1 & -1.87 & 0.11 & 1.81 &  $\ge 126$  & 20.0 &  $\le 0.6$ \\
SWIFT J0433.5-5846  & 0.103 & 44.4 & 2.50E-26 & 7.11E-26 & -0.45 &  RQU  & 2 & 8.5 & 44.6 & -1.98 & 0.65 & 1.45 &  $\ge 22$  & 23.2 &  $0.6$ \\
SWIFT J0445.0-2816  & 0.147 & 45 & 6.74E-23 & 1.16E-25 & 2.77 &  RL  & 2 & 9.11 & 45.2 & -2.05 & 0.64 & 1.44 &  $\ge 43$  & 22.0 &  $\le 0.5$ \\
SWIFT J0508.1+1727  & 0.0174 & 43.2 & 2.50E-26 & 9.37E-25 & -1.57 &  RQU  & 1 & 8.59 & 44.1 & -2.55 & 0.12 & 1.98 &  $\ge 46$  & 22.3 &  $3.8$ \\
SWIFT J0516.2-0009  & 0.0325 & 44.2 & 1.23E-25 & 2.20E-24 & -1.25 &  RQ  & 1 & 8.72 & 45.1 & -1.75 & 0.14 & 2.07 &  $\ge 292$  & 20.0 &  $1.7$ \\
SWIFT J0519.5-4545  & 0.0351 & 44 & 8.50E-22 & 6.23E-25 & 3.14 &  RL  & 2 & 8.7 & 44.6 & -2.21 & 0.28 & 1.79 &  $\ge 148$  & 20.5 &  $\le 0.9$ \\
SWIFT J0752.2+1937  & 0.117 & 44.7 & 2.49E-25 & 4.85E-25 & -0.29 &  RQ  & 1 & 9.36 & 45.6 & -1.89 & 0.12 & 1.28 &  $42$  & 22.5 &  $\le 1.6$ \\
SWIFT J0830.1+4154  & 0.126 & 44.7 & 4.70E-26 & 1.40E-25 & -0.47 &  RQ  & 1 & 8.56 & 45.1 & -1.56 & 0.4 & 1.8 &  $\ge 80$  & 21.1 &  $\le 9.1$ \\
SWIFT J0832.5+3703  & 0.0923 & 44.5 & 8.21E-26 & 2.79E-25 & -0.53 &  RQ  & 1 & 9.13 & 45.1 & -2.12 & 0.26 & 1.87 &  $\ge 142$  & 20.0 &  $\le 4.4$ \\
SWIFT J0839.6-1213  & 0.198 & 45.5 & 1.86E-23 & 2.31E-25 & 1.91 &  RL  & 1 & 8.86 & 45.8 & -1.21 & 0.5 & 2.03 &  $\ge 272$  & 20.0 &  $\le 4$ \\
SWIFT J0918.5+1618  & 0.0295 & 43.8 & 6.10E-26 & 2.76E-24 & -1.66 &  RQ  & 1 & 8.48 & 45.1 & -1.5 & 0.05 & 2.08 &  $\ge 261$  & 20.0 &  $2.9$ \\
SWIFT J0947.7+0726  & 0.0864 & 44.6 & 7.64E-23 & 2.68E-25 & 2.45 &  RL  & 1 & 8.9 & 45 & -1.96 & 0.39 & 1.77 &  $\ge 90$  & 20.0 &  $\le 0.6$ \\
SWIFT J1031.9-1418  & 0.0851 & 44.8 & 1.43E-25 & 9.88E-25 & -0.84 &  RQ  & 1 & 8.59 & 45.6 & -1.1 & 0.15 & 1.86 &  $154$  & 20.0 &  $\le 1.8$ \\
SWIFT J1052.8+1043  & 0.088 & 44.5 & 1.02E-25 & 3.24E-25 & -0.5 &  RQ  & 2 & 9.04 & 45.1 & -2.01 & 0.21 & 1.22 &  $\ge 35$  & 23.2 &  $\le 0.5$ \\
SWIFT J1115.3+5423  & 0.0703 & 44.4 & 6.60E-26 & 2.65E-25 & -0.6 &  RQ  & 2 & 8.58 & 44.8 & -1.84 & 0.39 & 1.69 &  $\ge 82$  & 22.8 &  $3.7$ \\
SWIFT J1138.9+2529B  & 0.219 & 44.9 & 8.29E-26 & 1.21E-25 & -0.16 &  RQ  & 2 & 8.87 & 45.6 & -1.4 & 0.21 & 1.56 &  $\ge 13$  & 23.1 &  $\le 0.8$ \\
SWIFT J1149.3+5307  & 0.0955 & 44.1 & 3.90E-26 & 6.19E-26 & -0.2 &  RQ  & 1 & 8.61 & 44.5 & -2.22 & 0.43 & 1.67 &  $47$  & 21.0 &  $\le 3.1$ \\
SWIFT J1158.9+4234  & 0.0314 & 43.5 & 4.67E-25 & 8.05E-25 & -0.24 &  RQ  & 2 & 8.6 & 44.6 & -2.1 & 0.09 & 1.75 &  $\ge 68$  & 23.9 &  $\le 1.1$ \\
SWIFT J1207.5+3355  & 0.079 & 44.2 & 4.90E-24 & 1.09E-25 & 1.65 &  RL  & 2 & 8.55 & 44.6 & -2.08 & 0.44 & 1.5 &  NA  & 22.7 &  $\le 0.8$ \\
SWIFT J1211.3-3935  & 0.0609 & 44.3 & 2.40E-25 & 3.78E-25 & -0.2 &  RQ  & 2 & 8.57 & 44.9 & -1.81 & 0.25 & 1.67 &  $\ge 69$  & 22.2 &  $\le 1.1$ \\
SWIFT J1222.4+7520  & 0.0703 & 44.3 & 9.21E-26 & 4.52E-25 & -0.69 &  RQ  & 1 & 8.61 & 45.1 & -1.64 & 0.18 & 2.05 &  $\ge 56$  & 20.0 &  $\le 1.3$ \\
SWIFT J1238.9-2720  & 0.0242 & 44.3 & 7.36E-25 & 1.27E-24 & -0.24 &  RQ  & 2 & 8.99 & 44.6 & -2.53 & 0.5 & 1.53 &  $152$  & 24.0 &  $0.11$ \\
SWIFT J1300.1+1635  & 0.0798 & 44.3 & 1.14E-25 & 9.41E-25 & -0.92 &  RQ  & 2 & 9.19 & 45.5 & -1.78 & 0.06 & 1.41 &  $\ge 79$  & 22.3 &  $\le 1.6$ \\
SWIFT J1309.2+1139  & 0.0252 & 44 & 2.50E-26 & 5.60E-25 & -1.35 &  RQ  & 2 & 8.54 & 44.2 & -2.4 & 0.53 & 1.21 &  $73$  & 23.7 &  $\le 0.3$ \\
SWIFT J1315.8+4420  & 0.0355 & 43.7 & 1.43E-25 & 1.06E-24 & -0.87 &  RQ  & 2 & 8.67 & 44.8 & -1.94 & 0.08 & 1.99 &  $\ge 68$  & 22.9 &  $0.7$ \\

\end{tabular}
\captionof{table}{Sample}
\end{sidewaystable}

\label{tbl_sample}
\restoregeometry

\clearpage

\newgeometry{left=6cm,bottom=1cm}

\begin{sidewaystable}
\scriptsize
\addtocounter{table}{-1}
\begin{tabular}{c c c c c c c c c c c c c c c c c} 
\hline

BAT NAME  &  $z$  &  $\log L_{14-195}$ & $F_{1.4}$    & $F_{\nu_{\rm W3}}$ &  $\log R$  \footnote{Radio loudness was calculated accordingly: $R = L_{1.4}/L_{\nu_{\rm W3}} \approx F_{1.4}/F_{\nu_{\rm W3}}$.}  & Radio class & Optical type &  $\log M_{\rm BH}/M_{\odot}$  &  $L_{\rm bol}$ \footnote{$L_{\rm bol}$ was calculated from $L_{\nu_{\rm W3}}$ using the relation $L_{\rm bol} = K_{W3} \nu_{W3} L_{\nu_{\rm W3}}$, where $K_{W3}= 8.4$ \citep{2006ApJS..166..470R}.} &  $\lambda_{\rm E}$  &  $L_{14-195}/L_{\rm bol}$  & $\Gamma_{14-195}$    &  $E_{\rm cut-off}$  &  $\log N_{H}$  &  $R_{\rm refl}$  \\
 \\ 
 & & $\rm erg\,s^{-1}$ & $\rm erg\,s^{-1} cm^{-2} Hz{-1}$ & $\rm erg\,s^{-1} cm^{-2} Hz{-1}$ & & & &  & $\rm erg\,s^{-1}$ & & & & keV & $\rm cm^{-2}$ &\\
 \hline\hline
SWIFT J1316.9-7155  & 0.0703 & 44.3 & 2.50E-26 & 3.32E-25 & -1.12 &  RQU  & 1 & 8.84 & 44.9 & -2 & 0.21 & 1.36 &  $50$  & 20.0 &  $\le 0.7$ \\
SWIFT J1331.2-2524  & 0.0261 & 43.3 & 1.26E-25 & 5.42E-25 & -0.63 &  RQ  & 1 & 8.56 & 44.3 & -2.4 & 0.1 & 1.52 &  NA  & 20.0 &  $\le 1.1$ \\
SWIFT J1341.2+3023  & 0.0407 & 43.9 & 4.25E-25 & 6.43E-25 & -0.18 &  RQ  & 1 & 8.54 & 44.7 & -1.91 & 0.15 & 1.65 &  $85$  & 23.5 &  $\le 0.5$ \\
SWIFT J1416.9-1158  & 0.0992 & 44.6 & 7.79E-26 & 2.01E-25 & -0.41 &  RQ  & 1 & 8.87 & 45 & -1.93 & 0.39 & 1.57 &  $56$  & 20.0 &  $\le 1.3$ \\
SWIFT J1427.5+1949  & 0.111 & 44.6 & 2.50E-26 & 3.75E-25 & -1.18 &  RQ  & 1 & 8.87 & 45.4 & -1.56 & 0.16 & 1.78 &  $\ge 60$  & 20.9 &  $\le 0.9$ \\
SWIFT J1429.2+0118  & 0.0865 & 44.5 & 2.70E-26 & 9.26E-25 & -1.54 &  RQ  & 1 & 8.67 & 45.6 & -1.19 & 0.08 & 2.18 &  NA  & 20.0 &  $\le 5.2$ \\
SWIFT J1508.8-0013  & 0.0544 & 44.2 & 2.29E-25 & 3.50E-25 & -0.18 &  RQ  & 1 & 8.61 & 44.7 & -1.98 & 0.26 & 1.79 &  $\ge 140$  & 20.3 &  $\le 0.3$ \\
SWIFT J1513.8-8125  & 0.0687 & 44.5 & 1.49E-25 & 5.16E-25 & -0.54 &  RQ  & 1 & 8.96 & 45.1 & -1.95 & 0.25 & 1.65 &  $\ge 98$  & 20.0 &  $\le 3$ \\
SWIFT J1521.8+0334  & 0.127 & 44.8 & 2.50E-26 & 1.02E-25 & -0.61 &  RQU  & 1 & 8.5 & 45 & -1.63 & 0.63 & 1.74 &  $\ge 55$  & 20.0 &  $\le 8.3$ \\
SWIFT J1542.0-1410  & 0.0967 & 44.4 & 2.50E-26 & 1.57E-25 & -0.8 &  RQU  & 1 & 8.54 & 44.9 & -1.74 & 0.32 & 1.74 &  $37$  & 22.2 &  $\ge 2.2$ \\
SWIFT J1547.5+2050  & 0.265 & 45.2 & 2.39E-23 & 1.70E-25 & 2.15 &  RL  & 1 & 9.12 & 45.9 & -1.31 & 0.21 & 1.61 &  $32$  & 20.0 &  NA \\
SWIFT J1548.5-1344  & 0.0244 & 43.8 & 3.00E-25 & 3.75E-24 & -1.1 &  RQ  & 1 & 8.64 & 45 & -1.7 & 0.06 & 1.67 &  $125$  & 22.0 &  $\le 0.9$ \\
SWIFT J1553.6+2347  & 0.118 & 44.5 & 6.29E-24 & 1.51E-25 & 1.62 &  RL  & 1 & 9.43 & 45.1 & -2.46 & 0.28 & 1.9 &  $\ge 24$  & 22.2 &  $\le 9.4$ \\
SWIFT J1607.2+4834  & 0.125 & 44.7 & 2.50E-26 & 1.14E-25 & -0.66 &  RQ  & 1 & 8.52 & 45 & -1.62 & 0.45 & 1.73 &  $\ge 151$  & 20.5 &  $\le 0.6$ \\
SWIFT J1614.0+6544  & 0.129 & 44.7 & 3.90E-26 & 8.81E-25 & -1.35 &  RQ  & 1 & 8.99 & 45.9 & -1.17 & 0.06 & 1.92 &  $\ge 43$  & 20.0 &  $\le 2.9$ \\
SWIFT J1617.8+3223  & 0.151 & 45.1 & 2.60E-23 & 1.74E-25 & 2.17 &  RL  & 1 & 8.64 & 45.4 & -1.37 & 0.55 & 1.79 &  $\ge 179$  & 22.3 &  $\le 0.6$ \\
SWIFT J1631.7+2353  & 0.0591 & 43.9 & 4.20E-26 & 2.02E-25 & -0.68 &  RQ  & 2 & 8.74 & 44.6 & -2.27 & 0.21 & 1.66 &  $43$  & 21.7 &  NA \\
SWIFT J1742.2+1833  & 0.186 & 45.1 & 1.14E-23 & 1.08E-25 & 2.02 &  RL  & 1 & 9.35 & 45.4 & -2.09 & 0.5 & 1.8 &  NA  & 20.0 &  $\le 4.4$ \\
SWIFT J1745.4+2906  & 0.111 & 44.8 & 1.30E-25 & 2.57E-25 & -0.3 &  RQ  & 1 & 8.82 & 45.3 & -1.67 & 0.36 & 1.73 &  $\ge 112$  & 20.0 &  $\le 4.1$ \\
SWIFT J1807.9+1124  & 0.0787 & 44.6 & 2.50E-26 & 1.34E-25 & -0.73 &  RQU  & 1 & 8.7 & 44.6 & -2.15 & 0.8 & 1.7 &  NA  & 21.5 &  $\le 0.4$ \\
SWIFT J1835.0+3240  & 0.0579 & 44.8 & 5.64E-23 & 1.01E-24 & 1.75 &  RL  & 1 & 8.9 & 45.2 & -1.76 & 0.34 & 2.07 &  NA  & 20.0 &  $\le 0.2$ \\
SWIFT J1842.0+7945  & 0.0561 & 44.9 & 1.16E-22 & 1.24E-24 & 1.97 &  RL  & 1 & 8.8 & 45.3 & -1.6 & 0.37 & 1.74 &  $166$  & 20.8 &  $\le 0.2$ \\
SWIFT J1947.3+4447  & 0.0528 & 44.2 & 4.70E-26 & 6.23E-25 & -1.12 &  RQ  & 2 & 9.01 & 45 & -2.16 & 0.19 & 1.53 &  $81$  & 22.8 &  $\le 0.3$ \\
SWIFT J1952.4+0237  & 0.0584 & 44.6 & 6.04E-23 & 8.01E-25 & 1.88 &  RL  & 2 & 8.96 & 45.2 & -1.91 & 0.26 & 1.69 &  $\ge 100$  & 23.7 &  $\le 0.2$ \\
SWIFT J1959.4+4044  & 0.0558 & 45 & 1.60E-20 & 1.31E-24 & 4.09 &  RL  & 2 & 9.4 & 45.3 & -2.17 & 0.47 & 1.48 &  $96$  & 23.4 &  $\le 0.8$ \\
SWIFT J2116.3+2512  & 0.153 & 44.8 & 2.50E-26 & 8.08E-26 & -0.51 &  RQU  & 1 & 8.92 & 45 & -1.97 & 0.5 & 1.84 &  $\ge 36$  & 20.0 &  $\le 3$ \\
SWIFT J2217.0+1413  & 0.0704 & 44.1 & 2.50E-26 & 6.92E-25 & -1.44 &  RQU  & 1 & 8.5 & 45.3 & -1.34 & 0.06 & 1.95 &  $\ge 25$  & 20.0 &  $\le 2.4$ \\
SWIFT J2223.9-0207  & 0.0601 & 44.7 & 5.79E-23 & 1.44E-24 & 1.6 &  RL  & 1 & 8.89 & 45.4 & -1.56 & 0.17 & 1.74 &  $129$  & 23.5 &  $\le 0.4$ \\
SWIFT J2246.0+3941  & 0.0811 & 44.8 & 1.06E-22 & 2.56E-25 & 2.62 &  RL  & 1 & 8.54 & 45 & -1.68 & 0.63 & 1.31 &  $96$  & 23.8 &  $1.5$ \\
SWIFT J2304.8-0843  & 0.0468 & 44.8 & 3.26E-25 & 1.19E-24 & -0.56 &  RQ  & 1 & 8.55 & 45.1 & -1.52 & 0.42 & 1.81 &  $320$  & 20.0 &  $1.2$ \\
SWIFT J2308.1+4014  & 0.0728 & 44.2 & 2.50E-26 & 8.48E-26 & -0.53 &  RQU  & 1 & 8.52 & 44.4 & -2.25 & 0.61 & 1.56 &  $104$  & 21.6 &  $\le 0.8$ \\
SWIFT J2318.9+0013  & 0.0293 & 44 & 2.44E-25 & 3.20E-24 & -1.12 &  RQ  & 1 & 8.61 & 45.1 & -1.57 & 0.07 & 1.88 &  $\ge 146$  & 20.0 &  $\le 1.2$ \\
SWIFT J2351.9-0109  & 0.174 & 45.1 & 1.61E-23 & 3.30E-25 & 1.69 &  RL  & 1 & 8.71 & 45.8 & -1.03 & 0.22 & 1.78 &  $\ge 158$  & 20.0 &  $\le 3.4$ \\
SWIFT J2359.3-6058  & 0.0963 & 44.5 & 2.54E-22 & 1.09E-25 & 3.37 &  RL  & 2 & 8.96 & 44.7 & -2.32 & 0.58 & 1.59 &  $\ge 49$  & 23.2 &  $\le 1.4$ \\

\end{tabular}
\captionof{table}{Sample Continued}
\end{sidewaystable}
\restoregeometry
\clearpage

\bsp	
\label{lastpage}
\end{document}